\documentclass[12 pt]{article}
 \usepackage{amsmath, amsthm, amssymb, bigints, bm}
 \usepackage{color, mathtools, graphics}
 \usepackage{setspace}
 \usepackage{ulem}
 \usepackage{url}
 \usepackage{subcaption}
 \usepackage[margin=1.0in]{geometry}
  \usepackage{tcolorbox}
  \usepackage{booktabs}
	\usepackage{array,tabularx,booktabs}
\usepackage{adjustbox}
\doublespace
\newcommand{\bfb} {\mbox{\boldmath ${\bm \beta}$}}

\title{Integrative Learning for Population of Dynamic Networks with Covariates}
\author{Suprateek Kundu\footnote{Corresponding author email: suprateek.kundu@emory.edu and address: 1518 Clifton Road, Atlanta, GA 30322, USA}, Jin Ming\footnote{equal contribution by SK and JM who are co-first authors}, Joe Nocera, and Keith M. McGregor}

\date{}
\begin{document}

\maketitle

{\noindent \bf Abstract:}  Although there is a rapidly growing literature on dynamic connectivity methods, the primary focus has been on separate network estimation for each individual, which fails to leverage common patterns of information. We propose novel graph-theoretic approaches for estimating a population of dynamic networks that are able to borrow information across multiple heterogeneous samples in an unsupervised manner and guided by covariate information. Specifically, we develop a Bayesian product mixture model that imposes independent mixture priors at each time scan and uses covariates to model the mixture weights, which results in time-varying clusters of samples designed to pool information. The computation is carried out using an efficient Expectation-Maximization algorithm. Extensive simulation studies illustrate sharp gains in recovering the true dynamic network  over existing dynamic connectivity methods. An analysis of fMRI block task data with behavioral interventions reveal sub-groups of individuals having similar dynamic connectivity, and identifies intervention-related dynamic network changes that are  concentrated in biologically interpretable brain regions. In contrast, existing dynamic connectivity approaches are able to detect minimal or no changes in connectivity over time, which seems biologically unrealistic and highlights the challenges resulting from the inability to systematically borrow information across samples.

     
\section{Introduction}
There has been a steady development of graph-theoretic approaches to compute dynamic functional connectivity (FC) that is fueled by an increasing agreement that the brain network does not remain constant across time and instead undergoes temporal changes resulting from endogenous and exogenous factors (Filippi et al., 2019).  For example, task-related imaging studies have shown that the brain networks will re-organize when the subjects undergo different modulations of the experimental tasks during the scanning session (Chang and Glover, 2010; Lukemire et. al, 2020). Similarly, dynamic FC has also been observed during  resting-state experiments
(Bullmore and Sporns, 2009). These, and other recent studies, have found increasing evidence of underlying neuronal bases for temporal variations in FC which is linked with changes in cognitive and disease states (Hutchinson et al., 2013).

Dynamic connectivity approaches involve time-varying correlations derived via graph-theoretic methods, and may be broadly classified into the following categories: (i)  change point  methods (Cribben et al., 2013; Kundu et al., 2018) that assume stable phases of connectivity interspersed with connectivity jumps at unknown locations, which results in  piecewise constant connectivity;  (ii)  Hidden Markov Models (HMMs)  involving fast transient networks that are reinforced or revisited over time, which have been applied to   electrophysiological data (Quinn et al., 2018) and more recently to fMRI data (Warnick et al., 2018); and (iii) sliding window approaches that enforce temporally smooth correlations (Chang and Glover, 2010; Monti et al., 2014) based on the biologically plausible assumption of slowly varying temporal correlations with gradual changes in connectivity. Although sliding window methods are arguably the most widely used, these approaches may be limited by practical issues such as the choice of the window length (Lindquist et. al, 2014). 

On the other hand, change point models and HMMs have the advantage of model parsimony by limiting the distinct number of parameters. However, the performance of these methods often depend on modeling assumptions, and temporal smoothness of connectivity estimates can not be typically ensured. More importantly, since most of these existing approaches typically rely on  single-subject data, they often face challenges in terms of detecting rapid changes in connectivity and may result in inaccurate estimates due to a limited information from a single individual. 

Essentially, almost the entirety of the existing dynamic connectivity literature has focused on data from single individuals, due to the fact that temporal changes in connectivity are expected to be subject-specific and may not be replicated across individuals. However, recent evidence suggests that combining information across individuals in a group provides more accurate estimates for connectivity  (Hindriks et al., 2016), which adheres to the commonly used statistical principle of data aggregation using multiple samples to obtain more robust estimates. Kundu et. al (2018) proposed a sub-sampling approach to compute time varying dynamic connectivity networks using multi-subject fMRI data, which resulted in considerable gains in  dynamic network estimation under limited heterogeneity across samples, compared to a single-subject analyses. Unfortunately, the variations across samples may not be restricted in many practical settings.  To our knowledge, there is a scarcity of carefully calibrated approaches for pooling information across heterogeneous samples in order to accurately estimate a population of (single-subject) dynamic networks. This is perhaps not surprising, given that there are considerable challenges involved in developing such methods. From a methodological perspective, it is not immediately clear how to effectively borrow information across individuals in a data-adaptive manner that also respects the inherent connectivity differences between heterogeneous samples. Similarly when estimating dynamic networks with $V$ brain regions  for $N$ individuals each having $T$ time scans, one encounters computational challenges in terms of computing $NT$ distinct $V\times V$ connectivity matrices, which is not straightforward for high-dimensional fMRI applications. 

In this article, our goal is to develop a fundamentally novel hierarchical Bayesian product mixture modeling (BPMM) approach incorporating covariates (MacEachern, 1999) for estimating a population of dynamic networks corresponding to heterogeneous multi-subject fMRI data.  The importance of using covariates to model {\it known} stationary networks has already been illustrated in recent literature (Zhang et al., 2019; Sun and Li, 2017), where the networks are specified in advance. These methods suggest a strong justification for incorporating demographic, clinical, and behavioral covariates when modeling dynamic networks in order to obtain more accurate and reliable estimates (Shi and Guo, 2016). Motivated by these existing studies, the proposed BPMM framework estimates {\it unknown dynamic} networks by leveraging covariate information in order to inform the clustering mechanism under the mixture model, which is better designed to tackle heterogeneity across samples that ultimately results in more accurate network estimation. Under the proposed model, subgroups of individuals with similar dynamic connectivity profiles are identified, where the subgroup memberships are also influenced by covariate profiles and change over time in an unsupervised manner that is designed to pool information in order to estimate the dynamic networks. Another appealing feature of the proposed BPMM approach is the ability to report cluster level network summaries that are more robust to noise and heterogeneity in the data. Since the proposed approach clusters samples independently at each time scan guided by covariate information, it is clearly distinct from HMM approaches that instead cluster transient brain states across time scans. To our knowledge, the proposed approach is one of the first to estimate a population of dynamic networks incorporating covariate knowledge by integrating heterogeneous multi-subject fMRI data,  which represents considerable advances.

\begin{figure}
    \centering
    \includegraphics[width=\linewidth,height=5.5 in]{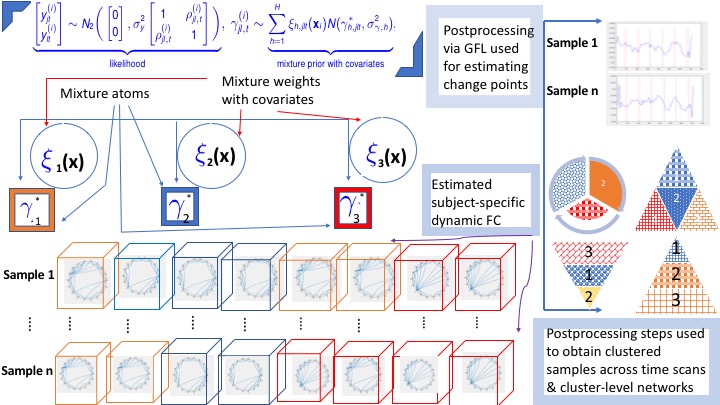}
    \caption{\small A schematic diagram illustrating the proposed dynamic pairwise correlation method. A mixture prior with $H=3$ components is used to model dynamic correlations, where the mixture weights are modeled using covariates. The resulting networks at each time scan for each sample are allocated to one of the $H$ clusters representing distinct network states that are represented by red, orange and blue cubes. Although the proposed method does not cluster transient states across time, the simplified representation in the Figure illustrates the similarity of brain states contained in identical colored cubes across the experimental session. Such temporal smoothness of the network is imposed via hierarchical fused lasso priors on the mixture atoms. Once, the dynamic FC is estimated, a post-processing step using K-means (Section 2.3) is applied to compute sub-groups of samples that exhibit similar dynamic connectivity patterns summarized across all time scans. The subgroups are represented by the circle, pyramid, triangle and inverted triangle shapes in the Figure and correspond to different modes of dynamic connectivity with  different number of brain states represented by different patterns within each shape. The connectivity change points for each individual, as well as at a cluster level, are computed via another post-processing step that employs a group fused lasso penalty (Section 2.4). The method reports both individual and cluster-level network features.}
    \label{fig:toy}
\end{figure}

In order to tackle the daunting task of estimating $NT$ connectivity matrices, each of dimension $V\times V$, the proposed approach employs dimension reduction by clustering samples under the mixture modeling framework that translates to considerable computational gains. In particular, the BPMM approach induces model parsimony by reducing the number of unique model parameters from $\frac{NT\times V(V-1)}{2}$ to $\frac{(\sum_{t=1}^T k_t)\times V(V-1)}{2}$, where $k_t(<<N)$ denotes the number of clusters at the $t$-time point that is determined in an unsupervised manner. Temporal smoothness in connectivity for each network is also ensured via additional hierarchical fused lasso priors on mixture atoms in the BPMM, which results in gradual changes in connectivity that is biologically meaningful. In scenarios where sharp connectivity changes are anticipated in certain localized time windows (due to changes in experimental design in a block task experiment, or other exogenous or endogenous factors), one may estimate these connectivity change points via a post-processing step that involves applying the total variation penalty (Bleakley and Vert, 2010) to the dynamic connectivity estimates under the proposed approach. Additional post-processing steps involving a K-means algorithm are also proposed to identify subgroups of individuals with similar dynamic connectivity patterns consolidated across time, which is particularly useful in terms of obtaining insights related to heterogeneity. Figure \ref{fig:toy} provides a visual illustration of the proposed approach.

 The proposed BPMM is developed for dynamic pairwise correlations as well as dynamic precision matrices, which provide complementary interpretations of dynamic connectivity. In particular, pairwise correlations encode connections between pairs of nodes without accounting for the effects of third party nodes, whereas partial correlations report association between nodes conditional on the effects of the remaining network nodes. While our goals do not involve assessing the merits of one approach over the other (see Smith et al., 2013 for a review), the proposed development is designed to provide users with an option to implement either approach as desired and suitable for respective applications. We develop an efficient Expectation-Maximization (EM) algorithm to  implement the dynamic pairwise correlation method separately for each edge, and another EM algorithm for dynamic precision matrix estimation that simultaneously involves all network nodes. We perform extensive simulations to evaluate the performance of the proposed method in contrast to existing approaches that involved a variety of dynamic network structures. The proposed methods were also used to investigate dynamic functional connectivity changes due to a high intensity, aerobic exercise 'spin' intervention when compared to a non-aerobic exercise, control intervention, which were administered to a heterogeneous group of sedentary adults who performed a fMRI block task experiment. Our goals are to provide connectivity insights that are complimentary to previous activation-based findings from the same study (McGregor et al., 2018), but involves analytic challenges due to the short duration of the fixation and task blocks that induce rapid connectivity changes which are usually  difficult to detect via existing methods.

The rest of the article is structured as follows. Section 2 develops the proposed approach for dynamic pairwise connectivity (denoted as integrative dynamic pairwise connectivity with covariates or idPAC) and dynamic precision matrices (denoted as integrative dynamic precision matrix with covariates or idPMAC), outlines a post-processing strategy for estimating network change points, as well as identifying clusters of samples with similar dynamic connectivity profiles. Section 3 develops a computationally efficient EM algorithm to implement the proposed approaches, and describes choices for tuning parameters. Section 4 reports results from extensive simulation studies, and Section 5 reports our analysis and results from the block-task fMRI experiment. Additional discussions are provided in Section 6. Throughout the article, we will use BPMM to denote the overall Bayesian product mixture modeling  framework used for developing the idPAC and idPMAC approaches, as appropriate.

\section{Methods}
In this section, we propose a novel approach for estimating a population of dynamic networks using heterogeneous multi-subject fMRI data with the same number of brain volumes across all individuals. 
For modeling purposes, we will assume that the demeaned fMRI measurements are normally distributed with zero mean (Kundu et al., 2018) at each time scan, and that pre-whitening steps have been performed to minimize temporal autocorrelations. We will fix some notations to begin with. Suppose fMRI data is collected for $T$ scans and $V$ nodes (voxels or regions of interest) for $N$ individuals. Denote the fMRI measurements across all the nodes at time point $t$ as ${\bf y}^{(i)}_t= (y^{(i)}_{1,t},\ldots,y^{(i)}_{V,t})'$, and denote the $V\times T$ matrix of fMRI measurements for the $i$-th individual as $Y^{(i)}$ that has the $t$-th column as ${\bf y}^{(i)}_t,i=1,\ldots,N$. Further, denote the vector of $q\times 1$ covariates as ${\bf x}_i$ for the $i$-th sample, and represent the collection of fMRI data matrices across all individuals as ${\bm Y}$.  

In what follows, we develop the idPAC method for pairwise correlations (Section 2.1) and idPMAC method involving partial correlations (Section 2.2), both of which involve a combination of likelihood terms and priors on the model parameters that are combined into a posterior distribution, which is used to estimate model parameters. The posterior distribution for parameter $\theta$ given data $Y$ is defined as $P(\theta| Y)=\frac{L(Y\mid \theta)\times \pi(\theta)}{P(Y)}$ using Bayes theorem, where $L(Y| \theta)$ denotes the data likelihood given the parameter value $\theta$, $\pi(\theta)$ represents the prior on $\theta$ under the Bayesian model, and $P(Y)=\int L(Y| \theta)\pi(\theta)d\theta$ is the marginal likelihood after integrating out all possible values of $\theta$. Full details of the posterior distributions for the idPAC models in (\ref{eq:base})-(\ref{eq:base_cov}) and the idPMAC model in (\ref{eq:base2})-(\ref{eq:base_cov2}) are provided in the Appendix.

\subsection{Dynamic Connectivity via Pair-wise Correlations}

Let the unknown dynamic functional connectivity (pairwise correlation) of individual $i$ be denoted as  $\bm{\rho^{(i)}}: = \{\rho^{(i)}_{jl,t}, j<l, j,l = 1 \dots V, t=1\dots T \}$, and the corresponding Fisher-transformed pairwise correlations be denoted as  $\gamma^{(i)}_{jl,t}={arctanh}(\rho^{(i)}_{jl,t})$.  We propose a Bayesian hierarchical approach that models the dynamic correlations for one edge at a time, using data from multiple individuals. We propose the following model for edge $(j,l)$, and $t=1,\ldots,T,$
\begin{eqnarray}
&&\begin{bmatrix}
y^{(i)}_{jt} \\ y^{(i)}_{lt}
\end{bmatrix} \sim N_2 \Bigg( \begin{bmatrix}
0 \\ 0 
\end{bmatrix} , \sigma_y^2 \begin{bmatrix}
1 & \rho^{(i)}_{jl,t} \\ \rho^{(i)}_{jl,t} & 1
\end{bmatrix} \Bigg) , 
\gamma^{(i)}_{jl,t} \sim \sum_{h=1}^H \xi_{h,jlt}({\bf x}_i) N(\gamma_{h,jlt}^*, \sigma^2_{\gamma,h}),\mbox{ } i=1,\ldots,N, \nonumber \\
&& \pi(\gamma_{h,jl1}^*, \ldots \gamma_{h,jlT}^*) \propto \exp(-\lambda \sum_{t=1}^{T-1}|\gamma^*_{h,jlt}-\gamma^*_{h,jl,t-1}|), \mbox{ } \sigma^{-2}_{\gamma,h}\sim Ga(a_\sigma,b_\sigma),\mbox{ } h=2,\ldots, H, 
\label{eq:base}
\end{eqnarray}
 where the Fisher-transformed correlations $\gamma^{(i)}_{jl,t}$ are modeled under a mixture of Gaussians prior having $H$ components denoted as $\gamma_{h,jlt}^*,h=1,\ldots,H,$  with the prior probability for the $h$-th mixture component denoted as $\xi_{h,jlt}({\bf x}_i)$ that depends on covariates, such that $\sum_{h=1}^H \xi_{h,jlt}({\bf x}_i)=1$ for all $t=1,\ldots,T$, $\sigma_y^2$ denotes the residual variance in the likelihood term, $\sigma^2_{\gamma,h}$ captures the (unknown) variability of the pairwise correlations under the mixture prior specification, $|\cdot|$ denotes the $L_1$ norm, and $N_v(\mu,\Sigma)$ denotes a multivariate Gaussian distribution with mean $\mu$ and $V\times V$ covariance matrix $\Sigma$. Under a hierarchical Bayesian specification,  $\sigma^{-2}_{\gamma,h}$ is estimated under the conjugate Gamma prior with shape and scale parameters $a_\sigma,b_\sigma,$ respectively. The mixture prior specifies that for any given time scan $t$, the functional connectivity for each individual can take values revolving around any one of the $H$ mixture atoms denoted by $(\gamma_{1,jlt}^*,\ldots,\gamma_{H,jlt}^*)$ with respective prior probabilities $(\xi_{1,jlt}({\bf x}_i),\ldots,\xi_{H,jlt}({\bf x}_i))$. These mixture probabilities and atoms are unknown and learnt adaptively from the data via posterior distributions under the proposed idPAC approach.
 
 {\noindent \emph{Modeling mixture atoms via fused lasso}}: The mixture atoms are modeled under a fused lasso prior in (\ref{eq:base}) that encourages temporal smoothness of pairwise correlations by assigning small prior probabilities for large changes in the values between consecutive time scans. Although temporal smoothness in correlations is encouraged, the Bayesian approach is still equipped to accommodate sharp jumps in connectivity that may arise due to changes in experimental design or other factors. Such connectivity jumps are detected using a post-processing step (see Section 2.4) applied to the estimated dynamic connectivity under the proposed model. 

{\noindent \uline{Modeling mixture weights via covariates:}} 
In order to effectively tackle heterogeneity, we incorporate supplementary covariate information when modeling the mixture weights under our mixture modeling framework in (\ref{eq:base}).  By incorporating covariate information, the model is designed to achieve more accurate identification of clusters, which then naturally translates to improved estimates for dynamic FC at the level of each individual. In particular,  we model $(\xi^{(i)}_{1,jl},\ldots,\xi^{(i)}_{H,jl})$ via a Multinomial Logistic regression (Engel, 1988)  as 
\begin{eqnarray}
\xi^{(i)}_{h,jlt}({\bf x}_i)= \frac{exp(\bm{x_i}^T \bfb_{h,jlt})}{1 + \sum_{h=1}^{H-1}exp(\bm{x_i}^T\bfb_{h,jlt})}, \mbox{ } \bfb_{h,jlt}\sim N(\bm{0},\Sigma_{\bfb}), \mbox{ } t=1,\ldots,T, h=1,\ldots,H-1, \label{eq:base_cov}
\end{eqnarray}
where $\small\bfb_{H,jlt}=0,t=1,\ldots,T,$ is fixed as the reference group, and $\small\bfb_{h,jlt},$ represent the vector of unknown regression coefficients that control the contribution of the covariates to the mixture probabilities for the $h$-th component ($h=1,\ldots,H-1$), in contrast to the $H$-th component. These regression coefficients are assigned a Gaussian prior with mean zero and prior covariance $\Sigma_\beta$ under a hierarchical Bayesian specification. A large value of these regression coefficients implies increased importance of the corresponding covariate with respect to modeling a particular edge under consideration, whereas $\small \bfb_{1,jlt}\approx \ldots \approx \bfb_{H-1,jlt}\approx 0$ for all $t=1,\ldots,T,$ indicates spurious covariates unrelated to the dynamic pairwise correlations.

Model (\ref{eq:base_cov}) suggests that the log-odds for each component $\xi^{(i)}_{h,jlt}({\bf x}_i)/\xi^{(i)}_{H,jlt}({\bf x}_i),h=1,H-1,$ can be expressed as a linear combination of covariates. When two or more samples have similar covariate information, the prior specification in (\ref{eq:base_cov}) will encourage similar mixture components to characterise the dynamic connectivity for all these samples that will result in analogous connectivity patterns. However the posterior distribution (that is used to derive parameter estimates) should be flexible enough to accurately estimate varying connectivity patterns between individuals even when they share similar covariate values, by leveraging information present in the data (as evident from extensive numerical studies in Section 4).

 {\noindent \uline{Role of clustering in tackling heterogeneity and pooling information:}}  Under model (\ref{eq:base}), each sample will be assigned to one of the $H$ clusters at each time scan in an unsupervised manner and guided by their covariate profiles in order to model the edge-level dynamic connectivity. Due to independent clustering at each time scan, these cluster configurations change over the experimental session in a data-adaptive manner to characterize connectivity fluctuations across individuals. Such time scan specific clusters represent subgroups of individuals with similar connectivity profiles over a subset of time scans, which are learnt by pooling information across all samples within a cluster. Here, it is important to note that model (\ref{eq:base}) does not impose identical dynamic connectivity across all time scans between multiple individuals (that is biologically unrealistic), but instead encourages common connectivity patterns within subgroups of samples for a subset of time points that are learnt in a data-adaptive manner. Hence, the proposed method is designed to result in more accurate estimation compared to a single subject analysis that is not equipped to pool information across samples or a group level analysis that does not account for within sample heterogeneity. We note that although the estimation is performed separately for each edge, the connectivity estimates across all edges are consolidated to obtain connectivity change point estimates (Section 2.3) or identify subgroups with common dynamic connectivity profiles (Section 2.4).

 \subsection{Dynamic Precision Matrix Estimation}
 
 We now propose a mixture model for dynamic precision matrix estimation that looks at the totality of all nodes in the network, in contrast to the edge-wise analysis in Section 2.1. While the proposed approach also uses a mixture modeling framework as in Section 2.1, the two methods are fundamentally distinct in the manner in which the mixture prior is specified and in terms of how the network edges are constructed and interpreted.  The proposed approach estimates the network by computing the $V\times V$ precision matrix involving $V(V-1)/2$ distinct partial correlations that are learnt by borrowing information across $V$ nodes at each time scan. The partial correlations measure interactions between pairs of regions after removing the influence of third party nodes, which is successful in filtering out spurious correlations. Hence a zero partial correlation between two nodes implies conditional independence. The proposed idPMAC approach enables one to report graph-theoretic network summary measures  that capture important patterns of network information transmission (Lukemire et al., 2020), which are otherwise difficult to report using pairwise correlations (Smith et al., 2012). 
 
Denote the $V\times V$ precision matrix over all nodes for the $i$-th individual at the $t$-th time point as $\small\Omega^{(i)}_t= \begin{bmatrix}
\omega^{(i)}_{t,11} &{\bm\omega}^{(i)}_{1,t} \\
{\bm\omega}^{(i)'}_{1,t} &\Omega^{(i)}_{11,t} \end{bmatrix}
$, and note that the partial correlation between nodes $k$ and $l$ is given directly as $-\omega_{kl}/\sqrt{\omega_{kk}\omega_{ll}}$ (ignoring the subject-specific and time-scan specific notations). We propose a Gaussian graphical model involving product mixture priors as:
\begin{eqnarray}
{\bf y}^{(i)}_t&\sim& N\Bigg[{\bf 0},\Omega^{(i)}_t
\Bigg], \mbox{ } {\bm\omega}^{(i)}_{v,t} \sim \sum_{h=1}^H \xi_{h,t}({\bf x}_i)N_{V-1}({\bm\omega}_{h,t}^*, \sigma^2_{\omega,h}I_{V-1}), \mbox{ } t=1,\ldots,T,i=1,\ldots,N, \nonumber \\
\omega^{(i)}_{t,vv} &\sim& E(\frac{\alpha}{2}), \mbox{ } \pi({\bm\omega}_{h,1}^*, \ldots, {\bm\omega}_{h,T}^*)\propto \exp(-\lambda \sum_{t=1}^{T-1}|{\bm\omega}^*_{h,t}-{\bm\omega}^*_{h,t-1}|), \mbox{ } h=1,\ldots, H,\label{eq:base2}
\end{eqnarray}
where $\Omega^{(i)}_t\in M^{+}_V$, the space of positive definite matrices, $E(\alpha)$ denotes the Exponential distribution with scale parameter $\alpha$, and  ${\bm\omega}^{(i)}_{v,t}$ denotes the vector of $(V-1)$ off-diagonal elements corresponding to the $v$-th row of $\Omega^{(i)}_t$ that are modeled using a mixture of multivariate Gaussians prior. Specifically, the dynamic connectivity at time scan $t$ is likely to be characterised via the $h$th mixture component with prior probability $\xi_{h,t}({\bf x}_i)$ depending on covariates, where the prior mean and variance for this unknown mixture component is given by  ${\bm\omega}^*_{h,t}$ and $\sigma^2_{\omega,h}$ respectively. The idPMAC approach in (\ref{eq:base2})-(\ref{eq:base_cov2}) specifies independent mixture priors on the set of all edges related to each node and at each time scan, which ensures symmetric and positive definite precision matrices that are necessary for obtaining valid partial correlation estimates. Full details for the computational steps are presented in Section 3.

{\noindent \emph{Modeling mixture atoms:}}
Under a hierarchical Bayesian specification, the mixture atoms or component-specific means ${\bm\omega}^*_{h,t}$ are themselves unknown and  modeled via a fused lasso prior, which encourages temporal homogeneity of partial correlations by assigning small prior probabilities for large changes in the values. We note that although the fused lasso prior encourages temporal smoothness in partial correlations, systematic changes in connectivity reflected by sharp jumps may be still identified via a post-processing step in Section 2.4. 


 {\noindent \underline{Modeling mixture weights via covariates:}} The node level mixture weights incorporating covariates are modeled via a Multinomial Logistic regression that is defined as:
\begin{eqnarray}
\xi^{(i)}_{h,t}({\bf x}_i)= \frac{e^{\bm{x_i}^T \bfb_{h,t}}}{1 + \sum_{h=1}^{H-1}e^{\bm{x_i}^T\bfb_{h,t}}}, \mbox{ } \bfb_{h,t}\sim N(\bm{0},\Sigma_{\bfb}),\mbox{ } i=1,\ldots,N, \label{eq:base_cov2}
\end{eqnarray}
where $\bfb_{h,t}$ refers to the unknown regression coefficients corresponding to time scan $t$ and mixture component $h$ that is assigned a Gaussian prior, and $\bfb_{H,t}=0,t=1\dots T,$ is set as the reference group. The prior in (\ref{eq:base2})-(\ref{eq:base_cov2}) encourages similar clustering configurations resulting in analogous time-varying partial correlations for individuals with similar covariate profiles. However in the presence of heterogeneity, the posterior distribution under the idPMAC method is able to identify divergent dynamic connectivity patterns even among individuals with similar covariate profiles (as evident from extensive numerical studies in Section 4).

 {\noindent \uline{Role of clustering in tackling heterogeneity and pooling information:}}  Under model  (\ref{eq:base2}), each column of the precision matrix is assigned to one of the $H$ clusters at each time scan in an unsupervised manner. Hence, the mixture modeling framework allows subsets of rows/columns of $\Omega^{(i)}_t$ to have the same values depending on their clustering allocation at each given time scan, which is an unique feature under the idPMAC approach that is not shared by the idPAC method. This feature results in robust estimates by pooling information across nodes and samples to estimate common partial correlations, and is a necessary dimension reduction step for scenarios involving large networks. For example, all weak or absent edges can be subsumed into one cluster which yields model parsimony. In addition, divergent connectivity patterns are captured via distinct time-varying clustering configurations across individuals as derived from the posterior distribution, which accommodates  heterogeneity. 
 Hence, the clustering mechanism under the idPMAC method not only enables dimension reduction, but also provides a desirable balance between leveraging common connectivity patterns within and across networks and addressing inherent network differences across individuals.

\subsection{Post-processing steps for sub-group detection}

In practical neuroimaging applications, it is often of interest to detect dissimilar modes of dynamic connectivity patterns that are embodied by distinct subgroups of individuals who also differ in terms of demographic or clinical characteristics, or other factors. For example in our fMRI task study, one of the objectives is to assess variations in dynamic connectivity with respect to subgroups of samples that were assigned different interventions, and who also had varying demographic characteristics. Instead of comparing network differences between pre-specified subgroups that are likely to contain individuals with  heterogeneous connectivity patterns, it is more appealing to develop a data-adaptive approach to identify subgroups that comprise individuals with homologous dynamic connectivity, and then examine connectivity variations across such subgroups and how these variations are related to intervention and other factors of interest.  When estimating these subgroups, we do not require identical dynamic connectivity patterns for all individuals within subgroups, but rather expect them to have limited network differences in terms of edge strengths and connectivity change points. An inherently appealing feature of subgroup detection is that is allows one to compute cluster level change points and other aggregate network features (see Section 2.4) which are more reproducible in the presence of noise and heterogeneity, compared to a single-subject analysis. Subgroup level network summaries may be particularly beneficial in certain scenarios such as fMRI block task experiments where it may be challenging for single-subject analyses to detect rapidly evolving network features induced via quick transitions between rest and task blocks within the  experimental design.


We propose an approach that consolidates the time-varying clusters of samples under the BPMM approach to detect subgroups which comprises samples with similar network-level dynamic connectivity patterns. In order to identify these subgroups, we first create a $N\times N$ similarity matrix that measures the propensity of each pair of samples to belong to the same cluster over the experimental session. This matrix is created by examining the proportion of time scans during which a pair of samples belonged to the same cluster across the experimental session, averaged across all edges. Once this similarity matrix has been computed, a K-means algorithm is applied to identify clusters of samples that exhibit similar dynamic connectivity patterns across the experimental session. The number of clusters $K$ is determined using some goodness of fit score such as the elbow method (Thorndike, 1953), or it is fixed as the maximum number of mixture components ($H$) under the BPMM approach. Finally, we note that the subgroup identification step is not strictly needed under the proposed BPMM framework for dynamic network estimation, but it is an optional analysis that can be used to identify cluster-level network features in certain scenarios of interest.

 \subsection{Post-processing steps for connectivity change point estimation}
 
 The estimated dynamic correlations in Sections 2.1-2.2 can be used to detect connectivity change points in scenarios involving sharp changes in the network during the session,  such as in fMRI task experiments. Our strategy involves computing change points for each individual network (a) at the edge level that captures localized changes; and (b)  at the global level that captures major disruptions in connectivity over the entire network. We compute the change points using the total variation  penalty (Bleakley and Vert, 2010) that was also used in CCPD approach by Kundu et. al (2018). However the proposed idPAC and idPMAC methods are distinct from the two-stage CCPD approach; the latter estimates connectivity change points based on empirical time-varying connectivity measures in the first stage, and then in the second stage, computes piecewise constant networks conditional on the estimated change points that represent connectivity jumps. In contrast, proposed idPAC and idPMAC methods pool information across samples in order to first estimate dynamic correlations that does not depend on change points and can vary continuously over time, and subsequently uses a post-processing step to compute connectivity change points without requiring piecewise constant connectivity assumptions. An appealing feature of the proposed mixture modeling framework guided by covariates is that it is more suitable for tackling divergent dynamic connectivity across samples,  in contrast to empirical correlations under the CCPD approach.

 Denote the vector of estimated (pairwise or partial) correlations over all edges for the $i$-th individual and at time scan $t$ as $\widehat{\bf r}^{(i)}_t \in \Re^{V(V-1)/2}, t=1,\ldots,T, i=1,\ldots,N$. Then the functional connectivity change points for the $i$-th individual may be estimated using connections across all edges via a total variation norm penalty that is defined as $\small ||{\bf u}^{(i)}_{t+1} - {\bf u}^{(i)}_{t} ||=\frac{1}{V(V-1)/2}\sqrt{\sum_{m=1}^{V(V-1)/2}(u^{(i)}_{t+1,m} - u^{(i)}_{t,m})^2}$. In particular, the following penalized criteria is used as in Kundu et al. (2018) for detecting network level connectivity change points:
\begin{eqnarray}
min_{{\bf u}\in \Re^{V(V-1)/2}} \sum_{t=1}^T|| {\widehat{\bf r}}^{(i)}_{t} - {\bf u}^{(i)}_t ||^2 + \lambda_u \sum_{t=1}^{T-1}||{\bf u}^{(i)}_{t+1} - {\bf u}^{(i)}_{t} ||, \label{eq:multiTV} 
\end{eqnarray}
where $\lambda_u$ represents the penalty parameter and ${\bf u}^{(i)}_{t}\in \Re^{p(p-1)/2}$ represents the piecewise constant approximation to the time series of correlations at time point $t$ for the $i$-th individual that also assumes the presence of an unknown number of connectivity jumps. The first term in (\ref{eq:multiTV}) measures the error between the observed correlations and the piece-wise constant connectivity, while the second term  controls the temporal smoothness of correlations for $V(V-1)/2$ edges. The increment $||{\bf u}^{(i)}_{t+1} - {\bf u}^{(i)}_{t} ||$ in the second term becomes negligible when the multivariate time series does not change significantly between times $t$ and $t+1$, but it takes large values corresponding to significant connectivity changes. The network change points computed via (\ref{eq:multiTV}) represent global changes functional connectivity resulting from a subset of edges that exhibit large connectivity changes. It is important to note that not all edges are expected to exhibit changes at these estimated change points. When it is of interest to compute edge-level connectivity change points, one can simply use criteria (\ref{eq:multiTV}) separately for each edge, so that the total variation term translates to the $L_1$ penalty. However, it is important to note that edge-level connectivity changes represent granular fluctuations that are typically more challenging to detect in the presence of noise in fMRI.

The number of change points is determined by the penalty parameter $\lambda_u$, with a smaller value  yielding a greater number of change points and vice-versa. Tibshirani and Wang (2007) proposed an estimate of $\lambda_u$ based on a pre-smoothed fit of a univariate time series using a lowess estimator (Becker et al., 1988). We adapt this approach for a multivariate time series to obtain an initial estimate for $\lambda_u$, and then propose post-processing steps to tune this estimate in order to obtain change points, as in the CCPD approach in Kundu et al. (2018). Full details for these steps are provided in Supplementary Materials. 

{\noindent \uline{Cluster-level connectivity change point estimation:}} 
For fMRI task experiments involving multiple subjects, subgroups of individuals are expected to share analogous dynamic connectivity patterns with limited variations across samples, as discussed in Section 2.3. The proposed total variation penalty norm in (\ref{eq:multiTV}) is equipped to leverage information across samples within a cluster for identifying cluster level change points, which reflect aggregated dynamic connectivity changes across all samples within a cluster at the global network level. These cluster level connectivity changes are obtained by aggregating the change points obtained via (\ref{eq:multiTV}) applied separately to each sample within the cluster, and then choosing those change points that show up repeatedly within the cluster. One can define a threshold such that all change points that appear with a high frequency (above the chosen threshold) across samples within the cluster are determined to represent cluster level change points (Kundu et al., 2018). We note that under the proposed method, it is entirely possible for individuals within a cluster to have unique connectivity changes in addition to the common cluster level change points, which reflect within sample heterogeneity. In our experience, this method typically works well in accurately recovering aggregated cluster-level connectivity changes, in certain scenarios such as block task experiments, or more generally in the presence of subgroups of individuals with similar dynamic connectivity patterns.


\section{Computational Details for Parameter Estimation}\label{sec:EM}
Although one can use Markov chain Monte Carlo (MCMC) to sample the parameters from the posterior distribution, we use a {\it maximum-a-posteriori} or MAP estimators for our purposes in this article that  bypasses the computational burden under a MCMC implementation. The MAP estimators are obtained by maximizing the posterior distribution for the model parameters and are derived via the Expectation-Maximization or EM algorithm. The EM algorithm is scalable to high-dimensional fMRI applications of interest that requires one to compute $N\times T$ distinct dynamic networks each involving $V\times V$ connectivity matrices.

\subsection{EM Algorithm for Pair-wise dynamic connectivity}
 
{\noindent \bf \uline{EM Algorithm:}} 
Denote the matrix containing the fMRI time series for the $l$th node as $Y_l=({\bf y}_{1,l},\ldots,{\bf y}_{T,l})$ where ${\bf y}_{t,l}=(y^{(1)}_{l,t},\ldots,y^{(N)}_{l,t})'$ represents the fMRI observations across all samples for node $l$ and time scan $t$.
Further, denote $\Delta_h$ as a latent indicator variable for the $h$th mixture component (that is not observed and is imputed in the proposed EM algorithm) and finally, denote by $\bm\Theta^{jl}$ the collection of all model parameters under the specification (\ref{eq:base})-(\ref{eq:base_cov}) corresponding to edge $(j,l)$. Note that under the proposed model (\ref{eq:base_cov}), one has an equivalent specification under the binary latent variables distributed as
$\small 
    (\Delta^{(i)}_{1,jlt},\ldots,\Delta^{(i)}_{H,jlt})\sim MN\big(1,(\xi_{1,jlt}({\bf x}_i;\bfb_{h,jlt}),\ldots,\xi_{H,jlt}({\bf x}_i;\bfb_{h,jlt})) \big),
$
where $MN(1;{\bf p}_0)$ denotes a multinomial distribution with probability vector ${\bf p}_0$, $B_{jlt}=({\bfb}_{1,jlt},\ldots,{\bfb}_{H-1,jlt} )$ and one can marginalize out $(\Delta^{(i)}_{1,jlt},\ldots,\Delta^{(i)}_{H,jlt})$ to recover the prior in (\ref{eq:base_cov}). The EM algorithm uses the augmented log-posterior derived in the Appendix involving the above latent mixture indicators, to computer MAP estimates for the model parameters by iteratively applying the Expectation (E) and Maximization (M) steps. The latent indicators $\small \{\Delta^{(i)}_{h,jlt},h=2,\ldots,H, t=1,\ldots,T,i=1,\ldots,N \}$ are imputed via the E-Step by using the posterior probability of $\gamma_{jl,t}^{(i)}$ taking values from the $h$-th mixture component, which is denoted by  $\psi_{h,jlt}^{(i)}=Pr(\Delta^{(i)}_{h,jlt} = 1\mid-)$ and updated as:

{\noindent \bf E-step}: Compute the posterior expectation for the latent cluster membership indicators as 
$\hat{\psi}_{h,jlt}^{(i)} = \big\{\sum_{r=1}^H \xi_{r,jlt}({\bf x}_i;\bfb_{h,jlt}) \phi(\gamma_{jl,t}^{(i)} \mid \gamma_{r,jlt}^*, \sigma^2_{\gamma,h})\big\}^{-1}\big\{ \xi_{h,jlt}({\bf x}_i;\bfb_{h,jlt}) \times \phi(\gamma_{jl,t}^{(i)} \mid \gamma_{h,jlt}^*, \sigma^2_{\gamma,h})\big\}$, where 
$\phi({\gamma_{jl,t}^{(i)}}\mid \gamma^*,\sigma^2_\gamma)$ denotes the normal density with mean $\gamma^*$ and variance   $\sigma_{\gamma}^2$. 

The remaining parameters are updated via M-steps using closed form solutions except $\gamma^{(i)}_{jl,t}$ that is updated using Newton-Raphson steps. These M-steps comprise several mathematically involved derivations and are detailed in the Appendix. The E and M steps are repeated till convergence, which occurs when the absolute change in the log-posterior between successive iterations falls below a certain threshold (we use $10^{-4}$ in our implementation).


\subsection{EM Algorithm for Dynamic Precision Matrix Estimation}
Let us denote the collection of all the precision matrices as $\bm{\Theta}$, and ${\bf y}^{(i)'}_{t,-v}$ as the $(V-1)$-dimensional vector of fMRI measurements at time scan $t$ over all nodes except node $v$. The prior on the precision matrix can be expressed as $\pi(\Omega^{(i)}_t)=\prod_{v=1}^V\pi(\omega^{(i)}_{t,vv})\pi({\bm\omega}^{(i)}_{vt})$, with the corresponding prior distributions $\pi(\cdot)$ being defined in (\ref{eq:base2}). Denote by $|\cdot|_1$, the element-wise $L_1$ norm, denote $\kappa^{(i)}_{1,t}=\omega^{(i)}_{t,11}-{\bm\omega}^{(i)'}_{1,t}\Omega^{(i)-1}_{11,t}{\bm\omega}^{(i)}_{1,t}$ to represent the conditional variance corresponding to the fMRI measurements for the $v$th node given all other nodes, and let  $\omega^{(i)}_{t,vv}$ and ${\bm\omega}^{(i)}_{v,t}$ respectively denote the diagonal and the vector of off-diagonal elements of the $v$th row in $\Omega^{(i)}_t$. Moreover use 
$det(A)$ to denote the determinant of the matrix $A$, and  write $\small S^{(i)}_t={\bf y}^{(i)}_t{\bf y}^{(i)'}_t= \begin{bmatrix}
s^{(i)}_{t,11} &{\bf s}^{(i)}_{1,t} \\
{\bf s}^{(i)'}_{1,t} &S^{(i)}_{11,t} \end{bmatrix}$ as the matrix of cross-products of the response variable, where $s^{(i)}_{vv,t}$ and ${\bf s}^{(i)}_{v,t}$ denote the $v$-th diagonal element and the off-diagonal elements for the $v$-th row respectively. Introduce latent indicator variables $(\Delta^{(i)}_{1,vt},\ldots,\Delta^{(i)}_{H,vt})$ that follow a multinomial distribution with probability vector  $(\xi_{1,t}({\bf x}_i),\ldots,\xi_{H,t}({\bf x}_i))$ such that $\sum_{h=1}^H \xi_{h,t}({\bf x}_i)=1$.

 Denote by $\Omega^{(i)}_{vv,t}$, the $(V-1)\times(V-1)$ obtained by deleting the $v$-th row and column from $\Omega^{(i)}_t$. The EM algorithm uses an E step for the latent mixture indicators, as well as a Monte Carlo E step that samples from the posterior distribution in order to obtain estimates for the precision matrix. These steps are described below:

{\noindent \bf E-step for mixture component indicator}: For $v=1,\ldots,V,$ use the expression:
$\hat{\psi}_{h,vt}^{(i)} =\big\{\sum_{r=1}^H\xi_{r,t}\big({\bf x}_i;\bfb_{h,t}\big) \phi_{V-1}\big({\bm\omega}_{v,t}^{(i)} \mid {\bm\omega}_{r,t}^*, \sigma^2_{\omega,r}I_{V-1}\big)\big\}^{-1}\times \big\{\xi_{h,t}\big({\bf x}_i;\bfb_{h,t}\big) \phi_{V-1}\big({\bm\omega}_{v,t}^{(i)} \mid {\bm\omega}_{h,t}^*, \sigma^2_{\omega,h}I_{V-1}\big) \big\}
$, where 
$\phi_{V-1}(\cdot\mid {\bm\omega}^*,\Sigma)$ denotes the probability density function for the ($V-1$)-dimensional normal density with mean and variance as  $({\bm\omega}^*,\Sigma)$ respectively. 

{\noindent \bf Monte Carlo E-step for precision matrix:} We use an E-step to update the precision matrix that computes the posterior mean by averaging MCMC samples drawn from the posterior distribution, which is equivalent to a Monte Carlo EM method (Wei and Tanner, 1990). We use this Monte Carlo approximation for the conditional expectation since it provides a computationally efficient approach to sample positive definite precision matrices via closed form posterior distributions. The  posterior distribution for the precision off-diagonal elements are given as  $\pi(\hat{\bm\omega}^{(i)}_{vt}\mid -) \sim N\Bigg[V_{\omega_{vt}}\bigg ( \sum_{h=1}^H\frac{\Delta^{(i)}_{h,vt} \omega_{h,t}^*}{\sigma_{\omega,h}^2} + 2({\bf s}^{(i)}_{v,t})\bigg ), V_{\omega_{v,t}}\Bigg]$, 
where $V_{\omega_{vt}}=\bigg(\sigma^2_{\omega,h}I_{V-1}+(s^{(i)}_{vv,t}+\alpha)(\Omega^{(i)-1}_{vv,t}) +\sum_{h=1}^H \frac{\Delta^{(i)}_{h,vt}}{\sigma_{\omega,h}^2}\bigg )^{-1}$ is the posterior covariance. Moreover, writing $\omega^{(i)}_{t,vv}=\kappa^{(i)}_{v,t}+{\bm\omega}^{(i)'}_{v,t}\Omega^{(i)-1}_{vv,t}{\bm\omega}^{(i)}_{v,t}$, the diagonal precision matrix elements are updated via the posterior 
$\kappa_{vt}^{(i)} \sim GA(\frac{1}{2}+1,\frac{s_{vv,t}^{(i)}+\alpha}{2})$
where $\alpha$ is pre-specified. The above steps can be alternated to sample positive definite precision matrices as in Wang (2012), and we draw several MCMC samples and average over them to approximate the conditional expectation.

The remaining parameters are updated via closed form expressions under the M step, which involve mathematically involved derivations and are detailed in the Appendix. The algorithm iterates through the E and M steps until convergence.

 \subsection{Tuning Parameter Selection}
 Certain tuning parameters in the BPMM need to be selected properly or pre-specified, in order to ensure optimal performance. For both dynamic pair-wise correlations and precision matrix estimation, $\lambda$ is the tuning parameter used in fused lasso penalty for the mixture atoms that controls the temporal smoothness of the dynamic connectivity. We choose an optimal value for $\lambda$ over a pre-specified grid of values, as the value of the tuning parameter that minimizes the BIC score.  In model (\ref{eq:base}) for the dynamic pairwise correlation, the $\sigma_y^2$ is also pre-specified as the initial mean variance over all edges and across all samples. Moreover when updating covariate effects, $\Sigma_{\beta}$ is pre-fixed as a diagonal matrix with the diagonal terms as $1$, although it is possible to impose a hierarchical prior on $\Sigma_\beta$ and update it using the posterior distribution. Extensive simulation studies revealed that the proposed approach is not sensitive to the choices of $\Sigma_\beta$  as long as the variances are not chosen to be exceedingly small. Other hyper-parameters in the hierarchical Bayesian specification include $\alpha$ in the prior on the precision matrices (chosen as in Wang (2012)), and $a_\sigma=0.1,b_\sigma=1,$ that results in an uninformative prior on the mixture variance.
 
The number of mixture components $H$ also needs to be chosen appropriately. On the one hand, a large value of $H$ may be used to address inherent heterogeneity, but it will also increase the running time and may generate redundant clusters that overcompensates for the variations across samples. On the other hand, a small value of $H$ may restrict the approach to overlook connectivity variations across individuals, resulting in inaccurate estimates.  
 One may  use a data adaptive approach to select $H$ in certain scenarios where it is reasonable to assume that the dynamic connectivity can be approximated by piecewise constant connectivity. In such cases that potentially involve block task experiments (Kundu et al., 2018), one can evaluate criteria (\ref{eq:multiTV}) separately for each individual under different values of $H$, and fix the optimal choice as that which minimizes the average value of the criteria (\ref{eq:multiTV}) across all individuals. Based on extensive empirical studies, we noticed the need for larger values for $H$ when fitting the model for cases involving large number of nodes and samples.



\section{Numerical Experiments}

\subsection{Simulation set-up}

{\noindent \underline{Data generation:}}  We generate observations from Gaussian distributions with sparse and piecewise constant precision matrices that change at a finite set of change points. Moreover, the network change points are generated based on covariate information where individuals with identical covariates have partially overlapping connectivity change points. Broadly, we use the following few steps to generate the data, each of which is described in greater detail in the sequel: (i) generate a given number of change points for each subject using corresponding covariate information; (ii) conditional on the generated change points, piecewise constant networks are simulated such that the connectivity changes occur only at the given change points; (iii) conditional on the network for a given state phase, a corresponding positive definite precision matrix is generated for each time scan where non-zero off-diagonal elements represent edge strengths and zero off-diagonals represent absent edges; and (iv) the response variable for a given time point is generated from a Gaussian distribution having zero mean and the precision matrix in step (iii). Four clusters are created with 10 samples each, where the samples with each cluster have the same number of connectivity change points, common state phase specific networks and identical covariate values. However within each cluster, there are differences in locations of connectivity change points and the network edge strengths are free to vary across individuals even when they share the same network structure. All samples in the first two clusters have 3 connectivity change points each, whereas the samples in the other two clusters have 4 change points, out of a total of $T=300$ time scans.

Conditional on the change points in step (i), several types of networks are constructed for each state phase in step (ii) that include: (a) Erdos Renyi network where each edge can randomly appear with a fixed probability; (b) small-world network, where the mean geodesic distance between nodes are relatively small compared with the number of nodes and which mimics several practical brain network configurations; and (c) scale-free network that resembles a hub network  where the degree of network follows a power distribution. Given these networks, the corresponding precision matrix was generated in step (iii) by assigning zeros to off-diagonals for absent edges, and randomly generating edge weights from uniform [-1,1] for all important edges. To ensure the positive definiteness, the diagonal values of the precision matrix were rescaled by adding the sum of the absolute values of all elements in each row with one. Finally, the response variables were generated either (a) independently at each time point via a Gaussian graphical model, or (b) via a vector autoregressive (VAR) model where the response variables are autocorrelated across time. In both cases, sparse time-varying precision matrices having dimensions $V=40,100,$ were used.

We generated two binary features that resulted in four distinct covariate configurations, i.e. (0,0), (0,1), (1,0), (1,1), and all samples with identical covariates were allocated to the same cluster.  In addition, we also evaluated the performance of proposed method in the presence of spurious covariates that are not related to dynamic connectivity patterns. Specifically, we introduced anywhere between 1 to 8 spurious covariates for each sample (in addition to the two true covariates described earlier), which were randomly generated using uniform as well as from random normal distributions. We then investigated the performance of the proposed approach over varying number of spurious covariates.  While the proposed approach is expected to work best in practical experiments involving a carefully selected set of covariates that influence dynamic connectivity patterns, our goal was also to investigate the change in performance as the number of spurious covariates increase.

{\noindent \underline{Competing methods}:} We perform extensive simulation studies to evaluate the performance of the proposed approach, and compare the performance with (a) change point estimation approaches such as the CCPD (Kundu et al., 2018) that can estimate single subject connectivity using multi-subject data in the presence of limited heterogeneity, and the dynamic connectivity regression (DCR) approach for single subjects proposed in Cribben et al. (2013);  (b) an  empirical sliding window based approach (SD) and the model-based SINGLE (Monti et al., 2014) method that uses sliding window correlations; and (c) a covariate-naive version of the proposed approach using the methods in Sections 2.1 and 2.2 (denoted as BPMM-PC and BPMM-PR respectively) that employs a multinomial distribution to model the mixture weights without covariates. While methods in (a) and (c) are designed to report connectivity change points, we augmented the sliding window approaches in (b) using a post-processing step similar to (\ref{eq:multiTV}) to compute change points based on the estimated sliding window correlations. Moreover for the proposed approach, the data under the VAR case was prewhitened via an autoregressive integrated moving average (ARIMA) before fitting the proposed models. In particular, the `$auto.arima$' in $R$ was used to prewhiten the raw data, which yielded residuals that were subsequently used for analysis. We note that it was not possible to report results under SINGLE for $V=100$ due to an infeasible computational burden.


{\noindent \underline{Performance metrics:}}
We evaluate the performance of different approaches in terms of different metrics. First, we investigated the accuracy in recovering true connectivity change points at the network and edge level for each sample, using sensitivity (defined as the proportion of truly detected change points or true positives), as well as the number of falsely detected change points or false positives. In addition, the performance of the network connectivity change points at the cluster level was also evaluated by comparing the true connectivity change points for each sample within the cluster with the aggregated cluster level change points. We note that since there were variations in connectivity change points within each cluster, false positive change points are to be expected under any estimation approach; however our goal is to evaluate how well these false positives are controlled and the sensitivity in detecting true change points under different methods. In addition, we also evaluated accuracy in terms of estimating the strength of connections that is computed as a squared loss (MSE) between the estimated and the true edge-level pairwise correlations. The pairwise correlations corresponding to dynamic precision matrix approaches for computing MSE were obtained  by inverting the respective precision matrices.

In order to evaluate the accuracy in dynamic network estimation, we computed the F-1 score defined as $\small 2 (\mbox{Precision}\times \mbox{Recall})/(\mbox{Precision + Recall})$, where Precision=$\small TP/(TP+FP)$ is defined as the true positive rate, and Recall=$\small TP/(TP+FN)$ represents the sensitivity in estimating the edges in the network. Here, $TP, FP, FN,$ refer to the number of true positive, false positive, and false negative edges that are obtained via binary adjacency matrices derived by thresholding the estimated absolute partial correlations. We employed reasonable thresholds (0.05) that are commonly used in literature (Kundu et al., 2018). In contrast, it was not immediately clear how to choose such thresholds for pairwise correlations given the fact that they tend to be typically larger in magnitude and have greater variability. Hence, we did not report F-1 scores corresponding to pairwise correlations, although one could do so in principle by choosing suitable thresholds to obtain binary adjacency matrices. Finally, we also evaluated the clustering performance in terms of the clustering error (CE) and Variation of Information (VI).  CE (Patrikainen and Meila, 2006) is defined as the maximum overlap between the estimated clustering with the true clustering, whereas VI (Meil$\check{a}$, 2007) calculates the entropy associated with different clustering configurations.

\subsection{Results}
The performance in terms of recovering the true clusters of subjects is provided in Table \ref{tab:clus_acc}, in the presence of two covariates that are both related to the true connectivity changes. It is clear from the results that incorporating covariate information results in near perfect recovery of the clusters, in contrast to the covariate-naive version of the method. For $V=100$, the dynamic pairwise correlation approach seems to have a slightly higher accuracy in terms of cluster recovery compared to the dynamic precision matrix approach when data is generated from a VAR model. 
Table \ref{tab:cluster_cp} reports the accuracy in recovering the true network-level change points under the proposed approaches at the level of the estimated clusters, as per discussions in Section 2.4. In this case, both idPAC and idPMAC methods are shown to have near perfect recovery of the true network connectivity change points when data is generated under GGM, and high sensitivity when data is generated under VAR. Moreover when using data from a VAR model, the idPAC method has a comparable or higher sensitivity but also higher false positives for $V=100$ in terms of detecting connectivity change points at the cluster level, compared to the idPMAC method. We note that although all samples within a cluster had identical covariate information, the proposed approach was able to accommodate within cluster connectivity differences that is evident from low false positives and high sensitivity when estimating cluster level change points. Moreover as seen from Tables \ref{tab:dypc}-\ref{tab:dypm}, the accuracy in recovering cluster level connectivity change points is considerably higher than the corresponding results at the level of individual networks. These results indicate the usefulness of aggregating information when it is reasonable to assume the existence of subgroups of individuals who share some similar facets of dynamic connectivity.

\begin{table}[]
    \centering
    \begin{tabular}{l|cccc|cccc}
    \hline
    &\multicolumn{4}{c|}{idPAC} &\multicolumn{4}{c}{BPMM-PC}\\
    &\multicolumn{2}{c}{V=40}
    &\multicolumn{2}{c|}{V=100}    
    &\multicolumn{2}{c}{V=40}
    &\multicolumn{2}{c}{V=100}  \\
   &CE&VI&CE&VI&CE&VI&CE&VI  \\
   \hline
   GGM+Erdos-Renyi&0&0&0&0&0.64&1.93&0.62&2.19\\
   GGM+Small-world&0&0&0&0&0.57&1.92&0.71&2.23\\
   GGM+Scale-free&0&0&0&0&0.63&2.01&0.66&2.19\\
   VAR+Erdos-Renyi &0&0&0&0&0.61&1.93&0.67&1.97 \\
   VAR+Small-World &0&0&0&0&0.59&1.88&0.61&1.90\\
   VAR+Scale-Free &0&0&0&0&0.61&1.78&0.61&1.93\\
 \hline
    &\multicolumn{4}{c|}{idPMAC} &\multicolumn{4}{c}{BPMM-PR}\\
    &\multicolumn{2}{c}{V=40}
    &\multicolumn{2}{c|}{V=100}  
    &\multicolumn{2}{c}{V=40}
    &\multicolumn{2}{c}{V=100}  \\
     \hline
   GGM+Erdos-Renyi &0&0&0&0&0.43&1.41&0.54&1.59\\
   GGM+Small-world &0&0&0&0&0.41&1.41&0.51&1.68\\
   GGM+Scale-free &0&0&0&0&0.43&1.49&0.60&1.78\\
   VAR+Erdos-Renyi &0.08&0.25&0.04&0.17  & 0.54 &1.51 & 0.66 &1.88\\
   VAR+Small-World &0&0&0.03&0.14  &0.48 &1.47 &0.58&1.91 \\
   VAR+Scale-Free  &0&0&0.04 &0.11 &0.49 &1.42 & 0.63 &1.75 \\
 \hline
    \end{tabular}
    \caption{ Clustering performance under different network types. GGM implies that the Gaussian graphical model was used to generate temporally uncorrelated observations; VAR implies a vector autoregressive model that was used to generate temporally dependent observations. For the VAR case, the observations were pre-whitened before fitting the model. }
    \label{tab:clus_acc}
\end{table}

\begin{table}[]
    \centering
    \begin{tabular}{l|cccc|cccc}
    \hline
&\multicolumn{4}{c|}{idPAC}&\multicolumn{4}{c}{idPMAC} \\
    &\multicolumn{2}{c}{V=40}
    &\multicolumn{2}{c|}{V=100}   
    &\multicolumn{2}{c}{V=40}
    &\multicolumn{2}{c}{V=100}  \\
   &sens&FP&sens&FP&sens&FP&sens&FP\\
    \hline  
    GGM+Erdos-Renyi &1 &2.15 &0.99& 1.58      & 0.97& 3.94 &0.99 &3.18\\
     GGM+Small-world   & 0.97 & 2.11 &1& 1.59  & 0.99 &4.18 & 0.98 &3.17\\
    GGM+Scale-free  & 0.99 &2.09&1&1.37 & 1 & 3.91 & 0.97 & 3.09\\
    \hline
    VAR+Erdos-Renyi  & 0.91 &3.71 &0.88& 3.66  & 0.87& 3.47 &0.87 &2.89\\
     VAR+Small-world & 0.84 & 3.44 &0.8& 3.09  & 0.82 &3.45 & 0.81 &2.98\\
    VAR+Scale-free   & 0.88 &3.29&0.84&3.68    & 0.85 &3.3 & 0.81 & 3.01\\
    \end{tabular}
    \caption{Cluster-based network change point estimation under the proposed approaches, assuming that all samples within a particular cluster have the same number and similar location of change points, with a limited degree of heterogeneity in functional connectivity. If this assumption holds, then the cluster level network change point estimation provides greater accuracy compared to the estimated change points at the level of individuals as reported in subsequent Tables.}
    \label{tab:cluster_cp}
\end{table}

\begin{table}[]
    \centering
    \begin{tabular}{l|ccccc|ccccc}
    \hline
    {\bf Results for V=40}  & \multicolumn{2}{c}{Network CP} &\multicolumn{2}{c}{Edge CP}& MSE & \multicolumn{2}{c}{Network CP} &\multicolumn{2}{c}{Edge CP}& MSE \\
         &sens &FP &sens&FP &MSE  &sens &FP &sens&FP &MSE\\
       \hline
     &\multicolumn{5}{c}{BPMM-PC} &\multicolumn{5}{c}{idPAC}\\
    \hline
     GGM+Erdos-Renyi&0.91&7.31&0.50&1.12&{\bf 0.1}&{\bf 1}& 2.75&{\bf 0.92}&1.08&{\bf 0.09}\\
     GGM+Small-world&0.92&5.99&0.47&1.03&0.12&{\bf 0.98}& 2.77&{\bf 0.92}&1.01&{\bf 0.08}\\
     GGM+Scale-free&0.91&7.29&0.49&1.19&0.12&{\bf 1}& 2.81&{\bf 0.92}&1.1&{\bf 0.09}\\
         \hline
    &\multicolumn{5}{c}{SD+GFL}&\multicolumn{5}{c}{CCPD}\\
    \hline
    GGM+Erdos-Renyi&0.3&3.13&0.09&2.97&0.29 &0.92&$\bm{2.15}$&0.31&4.1&0.16\\
    GGM+Small-world&0.29&3.31&0.09&3.08&0.27&0.92&$\bm{2.18}$&0.29&4.17&0.21\\
    GGM+Scale-free&0.29&3.08&0.09&2.99&0.24&0.91&$\bm{2.33}$&0.29&4.09&0.19\\
    \hline
   \hline
   \hline
     &\multicolumn{5}{|c}{BPMM-PC} &\multicolumn{5}{c}{idPAC}\\
       \hline
       VAR+Erdos-Renyi &0.68&6.55&0.43&1.08&0.2&{\bf 0.84}&5.57&{\bf 0.80}&1.06&{\bf 0.12}\\
       VAR+Small-world &0.66 &5.97&0.47 &1.14 &0.19 &{\bf 0.77} &5.54 &{\bf 0.74} &1.12 & {\bf 0.09} \\
       VAR+Scale-free & 0.59 &5.51&0.39&1.02&0.17&{\bf 0.78}&5.29&{\bf 0.73} &1.06&{\bf 0.09}\\
  \hline
    &\multicolumn{5}{c}{SD+GFL}&\multicolumn{5}{c}{CCPD}\\
    \hline
 VAR+Erdos-Renyi&0.41&7.72&0.13&3.06&0.26&0.55&{\bf 1.12}&0.18&4.33&0.21\\
    VAR+Small-world& 0.56 & 6.29 &0.14 &2.98 &0.19 & 0.64&{\bf 1.36} &0.17 &3.47 &0.23\\
    VAR+Scale-free&0.42 &6.99 &0.17 &3.13 &0.22 & 0.58 &{\bf 1.27} &0.19 &3.29 &0.2 \\
     \toprule
    \toprule
   {\bf  Results for V=100}  & \multicolumn{2}{c}{Network CP} &\multicolumn{2}{c}{Edge CP}& MSE & \multicolumn{2}{c}{Network CP} &\multicolumn{2}{c}{Edge CP}& MSE \\
         &sens &FP &sens&FP &MSE  &sens &FP &sens&FP &MSE\\
       \hline
     &\multicolumn{5}{c}{BPMM-PC} &\multicolumn{5}{c}{idPAC}\\
    \hline
     GGM+Erdos-Renyi&0.92&4.77&0.51&1.31&0.11&{\bf 1}&2.31&{\bf 0.83}&{\bf 1.16}&0.09\\
     GGM+Small-world&0.91&4.69&0.49&1.33&0.1&{\bf 1}&2.37&{\bf 0.82}&{\bf 1.17}&0.09\\
     GGM+Scale-free&0.91&4.71&0.50&1.31&0.11&{\bf 1}&2.29&{\bf 0.83}&{\bf 1.16}&0.09\\
         \hline
     &\multicolumn{5}{c}{SD+GFL}&\multicolumn{5}{c}{CCPD}\\
    \hline
    GGM+Erdos-Renyi&0.3&3.13&0.09&2.97&0.29&0.9&{\bf 1.12}&0.29&4.6&0.18\\
    GGM+Small-world&0.29&3.31&0.09&3.08&0.27&0.91&{\bf 1.18}&0.25&4.2&0.17\\
    GGM+Scale-free&0.29&3.08&0.09&2.99&0.27&0.91&{\bf 1.02}&0.27&4.4&0.17\\
    \hline
    \hline
   \hline
     &\multicolumn{5}{|c}{BPMM-PC} &\multicolumn{5}{c}{idPAC}\\
       \hline
VAR+Erods-Renyi&0.66&5.97&0.51&1.07&0.14&{\bf 0.82}&5.88&{\bf 0.81}&1.04&{\bf 0.11}\\
       VAR+Small-world&0.59&6.03&0.41&1.02&0.14&{\bf 0.75}&5.44&{\bf 0.74}&1.05&{\bf 0.12} \\
       VAR+Scale-free& 0.62&5.49&0.44&0.99&0.15&{\bf 0.77}&5.51&{\bf0.71}&1.11&{\bf 0.13}\\
  \hline
    &\multicolumn{5}{c}{SD+GFL}&\multicolumn{5}{c}{CCPD}\\
    \hline
VAR+Erdos-Renyi&0.37&8.03&0.1&3.14&0.15&0.55&{\bf 1.09}&0.17&3.75&0.22\\
    VAR+Small-world& 0.44&7.51&0.16&2.71&0.16&0.66&{\bf 1.44}&0.19&3.41&0.19\\
    VAR+Scale-free& 0.36&7.72&0.18&2.88&0.18&0.59&{\bf 1.31}&0.17&3.44&0.19\\
        \end{tabular}
    \caption{Results under the dynamic pair-wise correlation approaches for network and edge-level connectivity change-point estimation (Edge CP) accuracy and network changepoint (Network CP) estimation accuracy for $V=40,100$. GGM and VAR correspond to data generated from Gaussian graphical models and vector autoregressive models. Significantly improved metrics among the four approaches corresponding to the GGM data and separately for the VAR data, are highlighted in bold font.}
    \label{tab:dypc}
\end{table}

\begin{table}[]
    \centering
    \setlength\tabcolsep{2.5pt}
    \begin{tabular}{l|cccccc|cccccc}
    \hline
    {\bf Results for V=40}   & \multicolumn{2}{c}{Network CP} &\multicolumn{2}{c}{Edge CP}& MSE&F1 & \multicolumn{2}{c}{Network CP} &\multicolumn{2}{c}{Edge CP}& MSE&F1 \\
         &sens &FP &sens&FP &MSE&F1  &sens &FP &sens&FP &MSE&F1\\
       \hline
     &\multicolumn{6}{|c}{BPMM-PM} &\multicolumn{6}{c}{idPMAC}\\
       \hline
  GGM+Erdos-Renyi&0.85&6.99&0.32&1.04&0.1& 0.79 &${\bf 1}$&{\bf 5.2}&{\bf 0.79}&{\bf 0.89}&0.08&{\bf 0.88}\\
  GGM+Small-world&0.88&7.14&0.33&1.16&0.08&0.77 &${\bf 1}$&{\bf 5.11}&{\bf 0.81}&{\bf 0.91}&0.08&{\bf 0.9}\\
  GGM+Scale-free&0.87&7.36&0.33&1.19&0.08& 0.71 &${\bf 0.97}$&{\bf 5.6}&{\bf 0.77}&{\bf 0.92}&0.07&{\bf 0.89}\\
    \hline
       &\multicolumn{6}{c}{DCR} &\multicolumn{6}{c}{SINGLE}\\
    \hline
   GGM+Erdos-Renyi   &0.22&16.15&0.41&9.39&0.27&0.59
   &0.35&6.49&0.1&2.84&0.08&0.71\\
   GGM+Small-world    &0.19&11.83&0.49&9.66&0.22&0.61
   &0.32&6.55&0.09&2.88&0.07&0.77\\
   GGM+Scale-free   &0.21&10.92&0.49&9.058&0.23&0.62
    &0.33&6.01&0.09&2.94&0.07&0.69\\
   \hline
   \hline
   \hline
     &\multicolumn{6}{|c}{BPMM-PM} &\multicolumn{6}{c}{idPMAC}\\
       \hline
       VAR+Erdos-Renyi& 0.66&{\bf 4.45} &0.29&{\bf 1.16}&0.10&0.77&{\bf 0.79}&4.81&{\bf 0.68}&1.22&0.09&{\bf 0.81}\\
        VAR+Small-world & 0.59&5.12&0.27&{\bf 1.03}&0.1&0.74&{\bf 0.78}&{\bf 4.99}&{\bf 0.69}&{\bf 1.04}&0.09&{\bf 0.79}\\
       VAR+Scale-free & 0.61&4.77&0.31&{\bf 1.04}&0.12&0.77&{\bf 0.76}&{\bf 4.64}&{\bf 0.71}&{\bf0.99}&{\bf 0.09}&{\bf 0.82}\\
  \hline
      &\multicolumn{6}{c}{DCR} &\multicolumn{6}{c}{SINGLE}\\
    \hline
    VAR+Erdos-Renyi  &0.22&9.83&0.4&3.35&0.24 &0.64
    &0.42&7.35&0.13&3.11&0.27 &0.66\\
     VAR+Small-world     &0.24&10.14&0.33&3.61&0.23&0.63 
     & 0.44&7.12 &0.17 &3.04 &0.26&0.62\\
    VAR+Scale-free &0.21 &9.98&0.32&3.61&0.22&0.59
    & 0.38 &6.77 &0.21 &3.36 &0.23&0.6\\
    \toprule
    \toprule
    {\bf Results for V=100}   & \multicolumn{2}{c}{Network CP} &\multicolumn{2}{c}{Edge CP}& MSE&F1 &  \multicolumn{2}{c}{Network CP} &\multicolumn{2}{c}{Edge CP}& MSE &F1\\
         &sens &FP &sens&FP &MSE&F1  &sens &FP &sens&FP &MSE&F1\\
       \hline
     &\multicolumn{6}{|c}{BPMM-PM} &\multicolumn{6}{c}{idPMAC}\\
       \hline
  GGM+Erdos-Renyi&{\bf 0.92}&6.83&0.28&1.09&0.08&0.83&{\bf 0.97}&{\bf 5.1}&{\bf 0.82}&{\bf 0.89}&0.08&{\bf 0.89}\\
  GGM+Small-world&{\bf 0.91}&6.98&0.31&1.19&0.09&0.81&{\bf 0.97}&{\bf 5.44}&{\bf 0.81}&{\bf 0.99}&0.07&{\bf 0.87}\\
  GGM+Scale-free&{\bf 0.92}&7.44&0.32&1.25&0.08&0.81&{\bf 0.96}&{\bf 5.6}&{\bf 0.79}&{\bf 0.94}&0.07&{\bf 0.87}\\
    \hline
              &\multicolumn{6}{c}{DCR} &\multicolumn{6}{c}{SINGLE}\\
    \hline
   GGM+Erdos-Renyi &0.33&16.14&0.41&9.39&0.22&0.63 &&&&&& \\
    GGM+Small-world 
   &0.31&15.88&0.4&9.66&0.27&0.59&&&NA&&&\\
     GGM+Scale-free
   &0.34&16.82&0.39&10.08&0.27&0.64&&&&&&\\
     \hline
  \hline
   \hline
    &\multicolumn{5}{c}{BPMM-PM} &\multicolumn{6}{c}{idPMAC}\\
       \hline
VAR+Erdos-Renyi&0.73&4.41&0.29&1.18&0.14&0.77&{\bf 0.88}&4.22&{\bf 0.63}&{\bf 1.09}&0.13&{\bf 0.82}\\
      VAR+Small-world &0.56& 5.22&0.22&{\bf 0.91}&0.11&0.78&{\bf 0.72}&{\bf 4.87}&{\bf 0.61}&1.09&0.1&{\bf 0.81}\\
       VAR+Scale-free & 0.59&5.13&0.29&1.03&0.11&0.78&{\bf0.77}&{\bf 4.49}&{\bf 0.65}&1.08&0.09&{\bf 0.81}\\
  \hline
    &\multicolumn{6}{c}{DCR} &\multicolumn{6}{c}{SINGLE}\\
    \hline
VAR+Erdos-Renyi&0.23&9.92&0.43&3.19&0.16&0.64&&&&&&\\
    VAR+Small-world&0.31&10.23&0.37&3.37&0.19&0.67 &&&NA&&&\\
    VAR+Scale-free& 0.25&10.23&0.38&3.61&0.18&0.65&&&&&&\\
    \end{tabular}
    \caption{Results under the dynamic precision matrix estimation approaches for network and edge-level connectivity change-point estimation (Edge CP) accuracy and network changepoint (Network CP) estimation accuracy  for $V=40,100$. GGM and VAR correspond to data generated from Gaussian graphical models and vector autoregressive models respectively. Significantly improved metrics among the four approaches corresponding to the GGM data and separately for the VAR data, are highlighted in bold font.}
    \label{tab:dypm}
\end{table}

\begin{figure}[h]
    \centering
    \includegraphics[width=\linewidth]{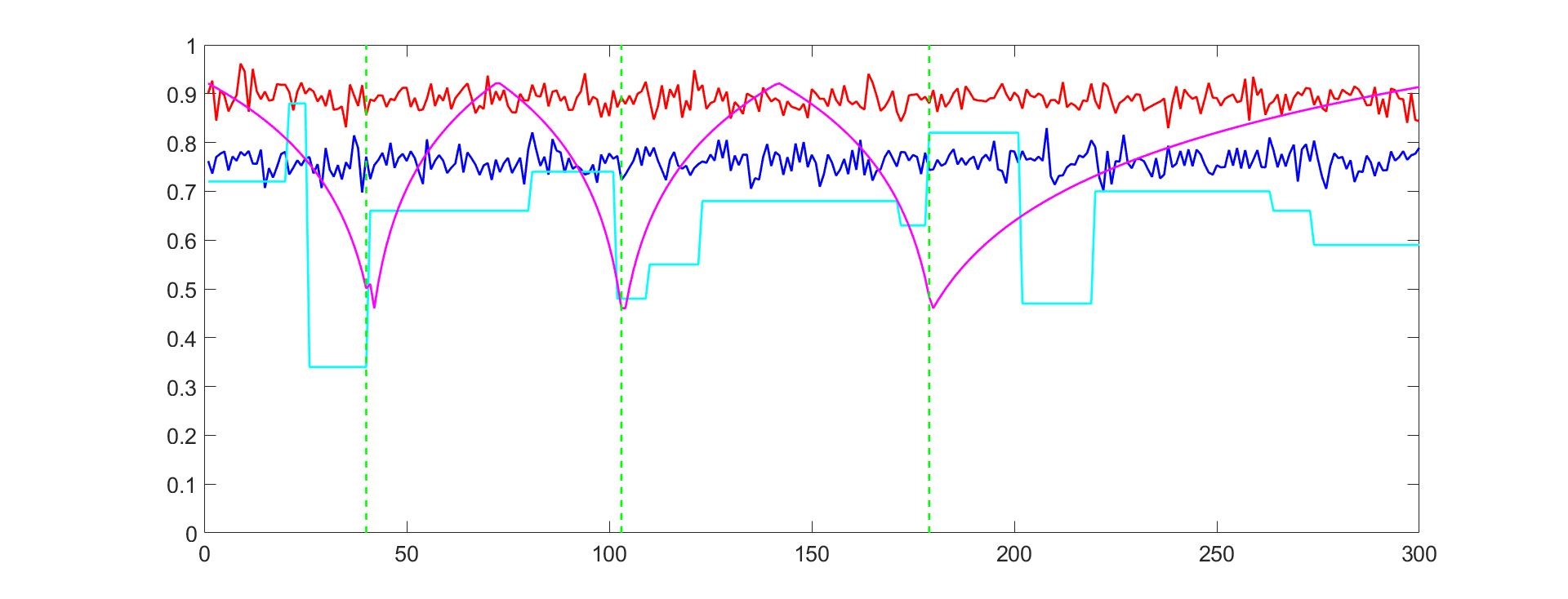}
    \caption{F1-score over time for one single subject under the case of dynamic partial correlation method. The vertical green lines are the true change points. Red line represents the proposed method with dynamic partial correlation (idPMAC), the cyan line represents the covariate-naive version (BPMM-PM), the blue line represents DCR, and the pink line represents SINGLE method.}
    \label{fig:f1score}
\end{figure}

Table \ref{tab:dypc} reports the performance under pair-wise correlation based approaches, i.e. idPAC, BPMM-PC, SD, and CCPD.  It is clear for the results that the proposed idPAC method has a near perfect sensitivity when data is generated under GGM, and a suitably high sensitivity under the VAR model, when estimating connectivity change points. The sensitivity for network and edge change point estimation, along with the MSE in estimating the pairwise correlations are significantly improved under idPAC compared to competing approaches in Table \ref{tab:dypc}. The CCPD method is shown to have the lowest false positives when estimating the network level change points, but otherwise has poor sensitivity for change point estimation and high MSE, which is potentially due to the assumption of piecewise constant connectivity. The approach based on sliding window correlations has the poorest performance across all the reported metrics, which illustrates their drawback in estimating dynamic connectivity.

Table \ref{tab:dypm} reports the performance under precision matrix based approaches, i.e. idPMAC,  BPMM-PR, SINGLE, and DCR. The results under the SINGLE method is not reported for $V=100$ due to infeasible computational burden. It is evident that the proposed idPMAC method has near-perfect or high sensitivity for detecting network level change points, corresponding to data generated under GGM and VAR models respectively. It also has a suitably high sensitivity for detecting edge level connectivity change points under both cases. Similarly, the MSE for edge strength estimation and the F-1 scores for network estimation accuracy are significantly improved under the proposed method in contrast to competing approaches. Figure \ref{fig:f1score} illustrates that the F-1 score over time under the proposed dynamic precision matrix method with covariates is almost always higher across almost all time scans compared to competing methods. Moreover the DCR and SINGLE method have the least impressive performance in terms of connectivity change point estimation, which also translates to poor dynamic network estimation (low F-1 scores).

Our results clearly illustrate the advantages of the proposed methods over existing  approaches that are not effective in leveraging information across samples. In addition,  Tables \ref{tab:dypc}-\ref{tab:dypm} also illustrate the gains of incorporating covariate information under the proposed idPAC and idPMAC approaches over the covariate naive BPMM counterparts. It is interesting to note that the covariate naive BPMM still fares better than existing dynamic connectivity methods that fail to pool information across samples in a systematic manner. We also note that while the presence of false positive (FP) connectivity change points are expected due to the heterogeneity across samples, the proposed approaches provide desirable control of FP even while pooling information across samples with varying networks. In fact, the FP under the proposed method are lower than all competing methods except CCPD, whose performance is otherwise less impressive in terms of significantly lower sensitivity for change point detection, and inferior network estimation as reflected by poor MSE and F-1 scores.

\begin{figure}[]
    \centering
    \includegraphics[width=1.05\linewidth, height=6.7in]{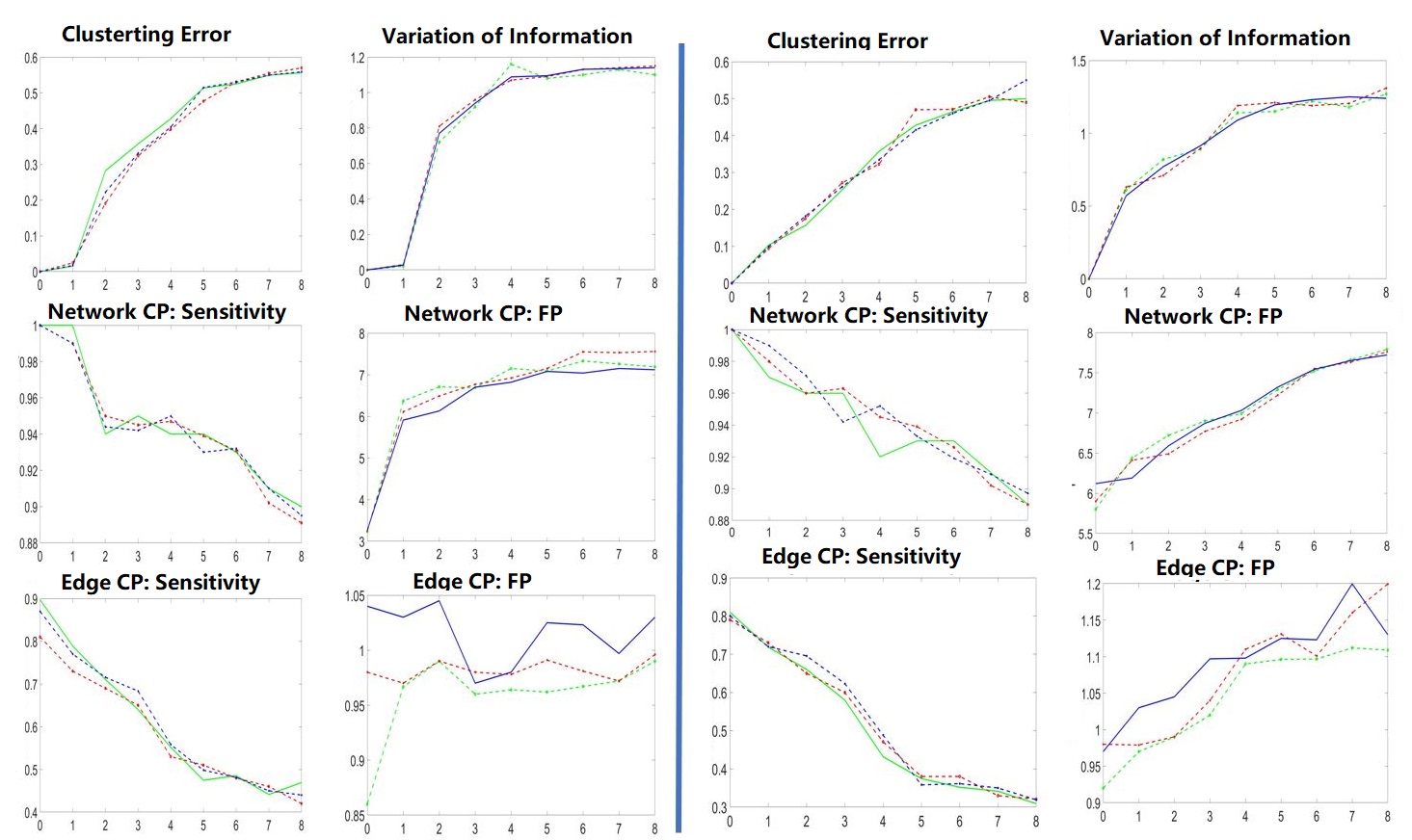}
    \caption{Performance of dynamic pairwise correlation (columns 1 and 2) and dynamic precision matrix (columns 3 and 4) methods under different number of spurious covariates represented by the X-axis. Lines with different color represent different network structure: Green (Erdos Renyi), Red (Small World), Blue (Scale Free). The top row provides the information of clustering performance (Clustering Error and Variation of Information), the middle row demonstrates the performance of network level change points estimation (sensitivity and number of False Positive estimations), and the performance of edge level change point estimation was provided in the bottle row.}
    \label{fig:spur_cov}
\end{figure}

When comparing the relative performance between idPAC and idPMAC methods, it is evident that the former has comparable or higher sensitivity but lower false positives in terms of estimating connectivity change points at the network level, when data is generated under a GGM. When data is generated under a VAR model, the idPAC method has higher sensitivity but also higher false positives compared to idPMAC, for estimating network connectivity change points. This is also true when estimating edge-level connectivity change points. In addition, since the idPMAC method  estimates all edges simultaneously, the mean squared error for estimating edge strengths is often lower compared to the idPAC method. Moreover when the number of spurious covariates is increased, both these approaches experience a drop in performance, as expected. However, while the rate of deterioration in terms of estimating connectivity change points is similar between the two methods (see second and third rows in Figure \ref{fig:spur_cov}), the dynamic precision matrix approach is more resilient to the presence of spurious covariates in terms of recovering the true clusters. This is evident from the top panels in Figure \ref{fig:spur_cov} that show a slower increase in the clustering error under the idPMAC method. 

The computation time for the proposed approaches are much faster compared to existing dynamic connectivity methods such as SINGLE, and comparable to the DCR approach proposed by Cribben et al. (2013). For example, it took the pairwise dynamic connectivity without covariates about 20 minutes to run for $V=20$, and the run time was around 26 minutes for this method with two covariates, with 40 individuals. Similarly, when $V=40$ and $T=300$, the average computation time is around 80 minutes with 40 subjects without covariates. The proposed method was scalable to $V=100$ and $T=300$, unlike the SINGLE approach whose  average computation time was around  6 hours. The total computation time under BPMM is expected to increase with $V,T,N,$ which is true for any method that computes dynamic connectivity at the level of each individual.

\section{Analysis of Task fMRI Data}
\subsection{Description of the study}
We analyze a block task data involving a semantic verbal fluency at Veterans Affairs Center for Visual and Neurocognitive Rehabilitation, Atlanta. In a 12-week randomized controlled trail, 33 elderly individuals (aged 60-80, 11 males, 22 females) were assigned to two intervention groups: spin aerobic exercise group (14 participants) and the non-aerobic exercise control group (19 participants). 
During the intervention, individuals belonging to the aerobic spin group were required to do 20-45 minutes of spin aerobic exercise three times a week, led by a qualified instructor. For control group, participants were asked to do the same amount of non-aerobic exercise per week, such as group balance and light muscle toning exercise.  A more detailed description of the data is available in Nocera et al. (2017).

For each participant, fMRI scans were conducted with 6 blocks of semantic verbal fluency (task) conditions with 8 scans, both pre- and post-intervention. The semantic verbal fluency task involved participants looking at different categories (e.g. ``colors") at the center of video screen and they were asked to generate and speak 8 different objects associated with that category (e.g. ``blue"). After task block, a rest block with 3-5 TRs would appear and participants were required to read the word ``rest" out loud. A total of 74 brain scans were acquired using a 3T Siemens Trio scanner with a whole-brain, 1-shot gradient EPI scan (240\time 240 mm FOV, 3.75 $\times$ 3.75 in-plane resolution, TR=5830ms, TA=1830ms, TE=25ms, flip angle (FA)=70). Analysis of Functional NeuroImages (AFNI) software and FMRIB Software Library (FSL) were used for pre-processing, as in Nocera et al. (2017). Slice-time corrections, linear trend removal, echo planar images alignment, and motion correction were performed as a part of the pre-processing pipeline.  We used 18 brain regions for analysis that were shown to be differentially activated between the two intervention groups as described in Nocera et al. (2017).  
These regions are listed in Table \ref{tab:ROI} and comprise more regions in the right hemisphere due to decreased activity in that hemisphere in the spin group following the intervention, as compared to the control group. We note that since these regions corresponded to group differences due to spin exercise, they can not be described as ``canonical" regions associated with semantic language function, which would also comprise some additional homologous regions in the left hemisphere. Since the purpose of the study was to investigate dynamic connectivity changes between brain regions due to the intervention, an analysis based on the selected 18 regions was undertaken instead of using canonical regions.

\begin{table}[]
    \centering
    \begin{tabular}{|l|l|c|l|}
    \hline
      ROI Number & Region name &Broadmann area & MNI coordinate\\
      \hline
  1 &R Cerebullum 1 & NA & (5,-62,-57)\\
  2 &R Inferior Temporal Gyrus &20 &(41,-27,-30)\\
  3 &R Angular Gyrus &39 &(44,-56,12)\\
  4 &R Middle Frontal Gyrus &10 &(23,56,-6)\\
  5 &R Middle Temporal Gyrus 1 &22 &(53,-12,-9)\\
  6 &L Precuneus 1 &7 &(-9,-74,57)\\
  7 &L Cingulate Gyrus &NA &(-9,-33,39)\\
  8 &R Precuneus &7 &(6,-80,48)\\
  9 &R Cerebellum 2 &NA &(35,-53,-27)\\
  10 &R Middle Temporal Gyrus 2 &21 &(60,-45,-6)\\
  11 &R Inferior Frontal Gyrus/precentral gyrus &44 &(59,9,9)\\
  12 &R Retrosplenial Area &30 &(9,-47,18)\\
  13 &R Supramarginal Gyrus &40 &(41,-36,33)\\
  14 &R Pars Triangularis/MFG &45 &(47,47,-9)\\
  15 &L Precuneus 2 &7 &(-6,-71,45)\\
  16 &L Cuneus &19 &(-15,-80,27)\\
  17 &L Superior Frontal Gyrus &6 &(-17,-18,69)\\
  18 &R Middle Temporal Gyrus 3 &22 &(60,-36,0)\\
  \hline
    \end{tabular}
    \caption{Summary of brain regions used for analysis. R and L are abbreviations for right and left respectively.}
    \label{tab:ROI}
\end{table}


\subsection{Analysis Outline}
We performed the analysis separately for the pre-intervention and post-intervention data, under both the dynamic pairwise correlations and dynamic precision matrix estimation methods.  We used age and gender as covariates for the pre-intervention dataset, while also using the type of intervention (spin or non-aerobic control) as an additional covariate for the post-intervention analysis. Our analysis is designed to: (i) investigate the clustering behavior and inspect how these clusters differ with respect to demographics and the intervention type; (ii) investigate the cluster-level network differences using network summary measures; (iii) estimate the connectivity change points and examine how well they align with the changes dictated by the block task experiment; (iv) infer nodes and edges in the network with significantly different connectivity patterns between pre- and post-intervention.

Objective (i) enables us to characterize homogeneous dynamic connectivity patterns corresponding to clusters of samples in terms of their demographic and clinical characteristics; aim (ii) will be instrumental in interpreting the cluster-level network differences that will shed light on network variations across transient network states; aim (iii) will provide insights regarding the effectiveness of the proposed approaches in terms of recovering connectivity jumps where these changes are influenced by, but often not fully aligned with, the changes in the block task experimental design (Hindriks et al., 2016; Kundu et al., 2018); and aim (iv) will inform investigators regarding dynamic connectivity differences that are associated with the type of intervention. For aim (ii), we were only able to report results under dynamic precision matrix estimation, since a graph theoretic framework is necessary to compute the network summary measures, which may not be feasible under a pairwise correlation analysis. 


\subsection{Results}

{\noindent \uline{Cluster analysis:}}
As seen from Table \ref{tab:clus_va}, the analysis under both idPAC and idPMAC methods yielded 5 clusters consolidated over all time scans (using the K-means algorithm described in Section 2.3), although the size of the clusters were more equitable under the idPAC method. The pre-intervention analysis yielded clusters that were largely homogeneous with respect to gender. These clusters were also reasonably well-separated with respect to age under the idPAC analysis, whereas the age of the participants within clusters were more diverse under the idPMAC analysis. The post-intervention analysis yielded more heterogeneous clusters with respect to both age and gender, with only one cluster comprising all males under both the idPAC and idPMAC analyses. This suggests a realignment of the dynamic connectivity after the intervention is administered, such that individuals with similar genders and age-groups have synchronous dynamic connectivity patterns pre-intervention as identified via subgroups, but the subgroups and their composition with respect to age and gender change post-intervention. Our post-intervention analysis also suggests that the variability across clusters under the idPAC method can be largely explained via the intervention type. 

\begin{table}[h]
    \centering
    \begin{tabular}{l|ccccc|ccccc}
    \hline
      Method & \multicolumn{5}{c}{idPAC} &\multicolumn{5}{c}{idPMAC}\\
      Cluster index   &1 &2 &3 &4 &5 &1 &2 &3 &4 &5\\
       \hline
      Cluster features & \multicolumn{5}{c}{Pre-intervention} & \multicolumn{5}{c}{Pre-intervention}\\\\
      \hline
  Size &8 &6 &8 &7 &4 &3 &5 &17 &6 &2\\
  \% of females &0 &100 &0 &14 &100 &0 &100 &0 &100 &0\\
  Age (mean) &72.2  &65.8  &64.7 &76.7 &67.7 &71.7 &69 &70.4 &66.8 &67\\
  Age(range) &69-73 &60-72 &60-68 &74-80 &66-69 &63-78 &62-80 &60-80 &60-72 &66-68\\
  CP(Task-Rest) &6 &3 &4 &4 &4 &4 &5 &5 &4 &3 \\
  CP(Rest-Task) &3 &5 &2 &3 &4 &4 &4 &2 &4 &3 \\
       \hline
              &\multicolumn{5}{c}{Post-intervention} &\multicolumn{5}{c}{Post-intervention}\\\\
    \hline
  Size &8 &4 &7 &11 &3 &3 &4 &9 &11 &6\\
  \% of females &63 &75 &0 &18 &33 &67 &100 &0 &9 &67\\
  Age (mean)   &67.3 &65 &65.1  &74.5. &73.7 &73.7  &69.3  &68.6 &73 &62.7 \\
  Age(range) &62-70 &60-71 &60-68 &71-80 &68-78 &67-80 &68-72 &63-78 &68-80 &60-66\\
  CP(Task-Rest) &5 &6 &4 &3 & 6 &3 &3 &5 &5 &5 \\
  CP(Rest-Task) &3 &5 &2 &2 &4 &2 &5 &2 &4 &2 \\
  Spin(\%) &0 &100 &100 &0 &100 &33 &0 &100 &9 &50\\
  \hline
    \end{tabular}
    \caption{Results for analysis of block task fMRI experiments. Size refers to the number of participants in each cluster, `CP(Task-Rest)' and `CP(Rest-Task)' denotes the cluster level connectivity change points that were detected within +/- 2 time scan of the change in experimental design from task to fixation, and from fixation to task, respectively. `Spin' refers to the percentage of individuals assigned the Spin intervention belonging to each cluster.}
    \label{tab:clus_va}
\end{table}

{\noindent \emph{Connectivity change point estimation:}}
Table \ref{tab:clus_va}  illustrates the cluster level connectivity change point estimation. We observed that under both the idPAC and idPMAC methods, the estimated change points were consistent with 4 or more (out of 6)  changes in experimental design when transitioning from task to rest, except one cluster where 3 of the connectivity change points aligned with the experimental design. These patterns were consistent in both the pre- and post-intervention analysis; however the number of connectivity change points that were strongly aligned with changes in the experimental design were (on average) greater in the post-intervention analysis compared to the pre-intervention analysis. This suggests a learning effect of the task that was reflected in terms of higher concordance between the connectivity change points and the experimental design post-intervention. On the other hand, the cluster-level estimation of change points when transitioning from fixation to task was (on average) less aligned with the experimental design compared to the change points when transitioning from task to fixation, as seen in Table \ref{tab:clus_va}. This is somewhat expected since there were only 3-5 time scans in each fixation block, which made it extremely challenging to detect connectivity changes when transitioning from fixation to task. However, the proposed approach was still able to detect at least two, and often 3 or more connectivity change points (out of 6) aligned with the experimental design that suggests a reasonable concordance between connectivity jumps and experimental transitions from fixation to task.



In contrast, the CCPD approach detected at most one or two connectivity change points, while the DCR method was not able to detect connectivity change points at all, which makes these results appear biologically impractical given the nature of the block task experiment. Although the changes in connectivity are not expected to be fully aligned with changes in the experimental design (Hindriks et al., 2016), one expects a certain degree of synchronicity between the two. Our results indicate that this is not captured at all via existing change point methods especially when there are rapidly occurring transitions in the experimental design, which  highlights their limitations. 
Hence, our analysis clearly illustrates the advantages of pooling information across heterogeneous samples and incorporating covariate knowledge via a mixture modeling framework, which is simply not possible using existing approaches that rely on information from single subjects as in DCR, or that use empirical methods to pool information across individuals as in CCPD. 

{\noindent \emph{Cluster level network differences:}} In order to investigate the differences between the networks corresponding to the different clusters, we examined variations in dynamic network metrics that capture  modes of information transmission in the brain. These network metrics include the characteristic path length (CPL) that measures the length of connections between nodes, and the mean clustering coefficient (MCC) that measures the clustering tendency averaged over all network nodes. Using permutation testing, we examined p-values to evaluate which pairs of clusters exhibited significantly different network summary measures. 
None of the clusters had significantly different CPL values in the pre-intervention analysis, but several pairs of clusters exhibited significant CPL differences post-intervention. The CPL differences were particularly pronounced between the first and remaining clusters, as well as the last and remaining clusters in the post-intervention analysis. These two clusters also demonstrated the highest within cluster variability in CPL values amongst all clusters. Moreover, the number of pairs of clusters with  significantly different MCC values increased from the pre-intervention to post-intervention analysis, with 8 out of 10 pairs of post-intervention clusters reporting significantly different MCC values compared to at least one other cluster. Hence, our results suggest greater variability in network organization between clusters in the post-intervention analysis compared to pre-intervention, which potentially reflects greater network heterogeneity after the 12 week intervention was administered.

\begin{figure}
    \centering
    \includegraphics[width=0.8\linewidth, height=3.7in]{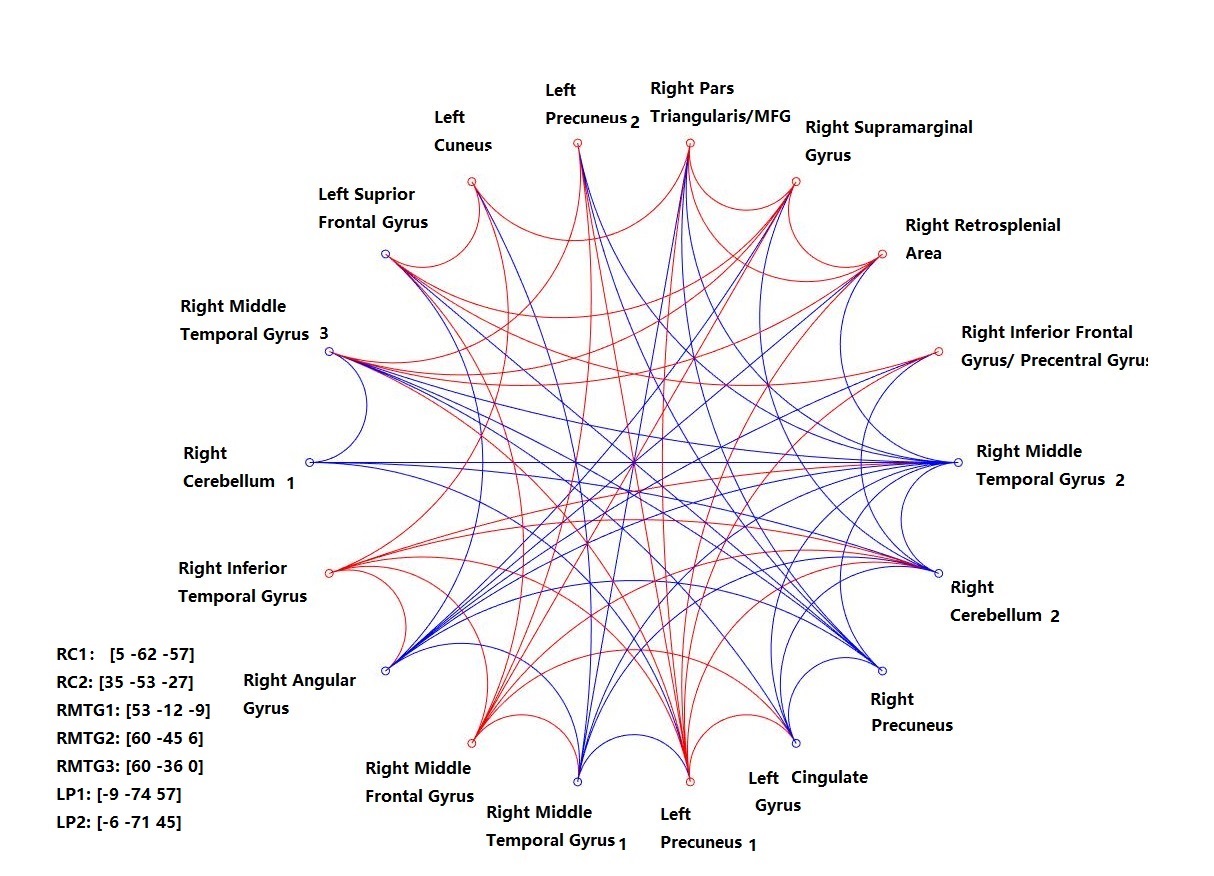}
    \includegraphics[width=0.8\linewidth,height=3.7in]{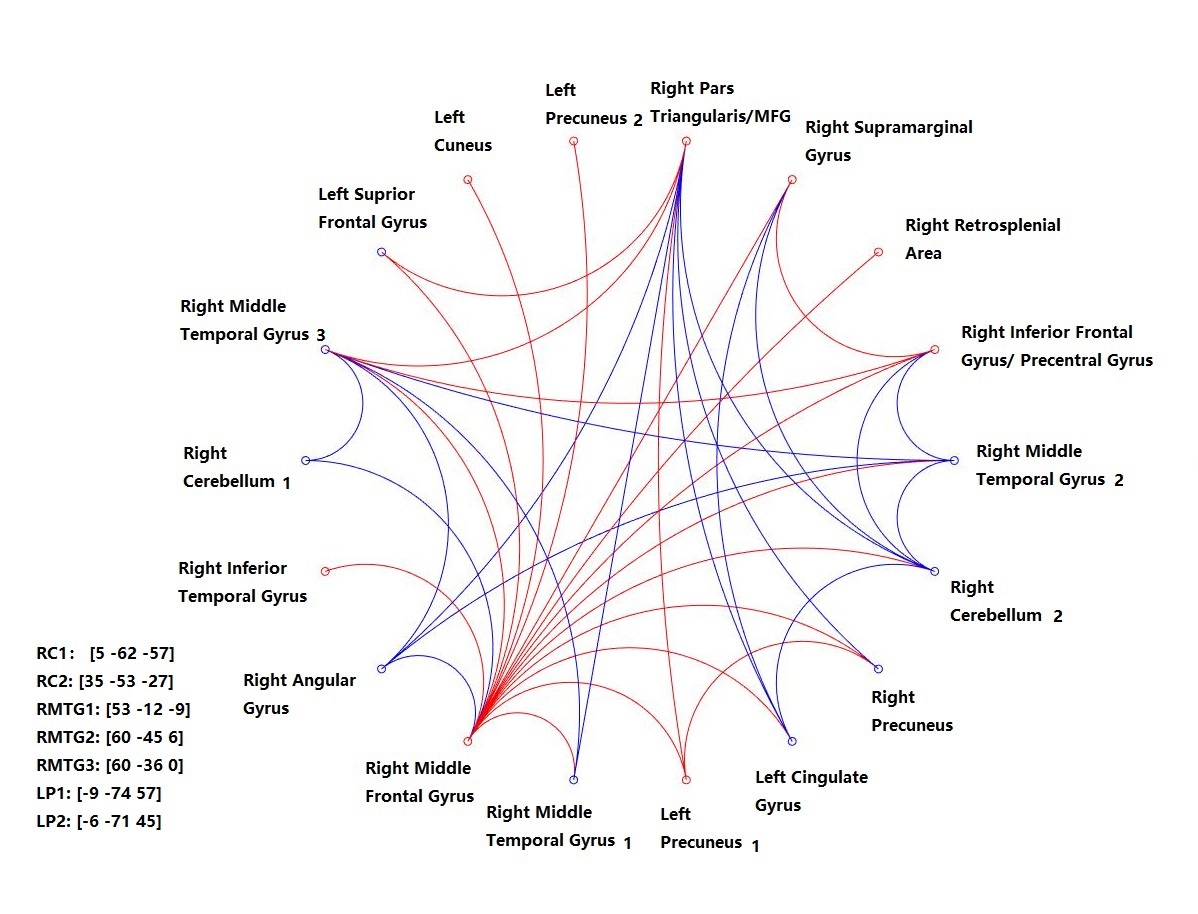}
    \caption{Circle plots for the edges that are significantly different pre- and post-intervention in spin group but not in the control group. The top and bottom panel correspond to the results under dynamic pairwise correlation and dynamic precision matrix estimation incorporating covariates, respectively. Red and blue lines correspond to lower or higher edge strengths in the pre-intervention network compared to post-intervention. RC1 and RC2 refer to the two brain regions in the right cerebellum; RMTG1-RMTG3 refer to the three brain regions in the right middle temporal gyrus; and LP1-LP2 refer to the two regions in the left precuneus. The MNI coordinates for these regions are provided in the Figure legend.}
    \label{fig:pre-post}
\end{figure}

{\noindent \emph{Network differences pre- and post-intervention:}} We applied paired t-test with multiplicity adjustment in order to infer which edges were significantly different between pre- and post-intervention at 5\% level of significance, along with identifying which network nodes contained the greatest number of differential edges. Since the magnitude of the pairwise correlations and the corresponding edge strength differences were higher, we discovered higher number of edges with differential edge strengths under the idPAC analysis. For both the idPAC and idPMAC methods, the bulk of the pre- vs post-intervention connectivity differences were concentrated in individuals in the spin group exclusively that were not present in the control group. We obtained 57 significantly different edges under the idPAC analysis, and 38 significantly different edges under the idPMAC analysis, which were exclusive to the spin group - see Figure \ref{fig:pre-post}. In contrast, the number of significantly different edges between the pre- and post-intervention networks under the idPAC analysis were 20 corresponding to both the spin and control groups, and 7 corresponding to the control group only. Moreover the idPMAC analysis did not produce any significant edge level differences between the pre- and post-intervention networks corresponding to both the intervention groups as well as for the control group only. Our results suggest a considerably strong realignment in dynamic  connectivity after the 12-week intervention that were exclusive to the spin group, compared to negligible changes in the control group.

The changes between the pre-vs post intervention networks that occurred exclusively in the spin group under idPAC analysis were concentrated in the following brain regions: Right Angular Gyrus(8 edges), Left Precuneus(10 edges), Right Cerebellum(9 edges), Right Middle Temporal Gyrus(11 edges), and Right Middle Temporal Gyrus(8 edges). Similarly the following brain regions had the highest number of differential edges pre- vs post-intervention under the idPMAC analysis: Right Middle Frontal Gyrus(16 edges), Right Cerebellum(6 edges), Right Pars Triangularis/MFG(8 edges), and Right Middle Temporal Gyrus(7 edges). Two nodes, Right Cerebellum and Right Middle Temporal Gyrus had a large number of significantly differential edges under both idPAC and idPMAC analyses, while the right middle frontal gyrus had, by far, the largest number of differential edges (16) under the dynamic precision matrix analysis. In addition, we also observe that more nodes in right hemisphere of the brain have significantly differential connectivity, which is to be expected since the majority of the 18 brain regions being investigated lie in the right hemisphere. 

 The large number of differential connections with respect to the right cerebellum is believed to be attributable to the generation of internal models or context specific properties of an object (Moberget et al., 2014), and preferential activation during a semantic challenge (D'Mello et al., 2017). The connectivity between the right cerebellum and inferior frontal regions has been noted in earlier studies (Balsters et al., 2013), with the inferior frontal regions being responsible for ordering language and codifying the motor output for syntax (Balsters et al., 2013). Moreover, the differential connectivity in the right middle temporal gyrus is along the lines of earlier findings that illustrated the role of the left temporal gyrus as a hub for integration of sensory input into a transformation to semantic forms (Davey et al., 2016), and the corresponding connectivity differences in the right middle temporal gyrus may be attributable to a shift in laterality of involvement (Lacombe et al. 2015) due to aging. Finally, the large number of differential edges corresponding to the right middle frontal gyrus is potentially associated with semantic priming in older adults (Laufer et al., 2011). Given that this region is associated with executive function (Wang et al., 2019; Jolles et al., 2013) and is well characterized as being involved in working memory tasks, it is likely for connectivity differences to be focused on this region since the semantic task requires a  continuous reference to working memory.

\section{Discussion}

In this article, we developed a novel approach that accurately estimates a population of subject-level dynamic networks by pooling information across multiple subjects in an unsupervised manner under a mixture modeling framework using covariates. The proposed approach, which is one of the first of its kind in dynamic connectivity literature, results in significant gains in dynamic network estimation accuracy, as illustrated via extensive numerical studies. The gains under the proposed method become particularly appealing compared to existing approaches in the presence of rapid transitions in connectivity as evident from our fMRI block task analysis. The proposed approach works best in fMRI task experiments involving a group of heterogeneous individuals executing the same task protocols, and in the presence of a carefully chosen set of covariates that are related to the dynamic network. 

We also illustrate the robust performance of the proposed approach in the presence of a limited number of covariates that are not related to changes in connectivity, although the performance deteriorates as the number of spurious covariates increase. In the presence of a large number of features that may not be necessarily related to dynamic connectivity, one can perform a screening step to exclude unimportant predictors from the analysis. This step will involve examining the associations between each covariate and the dynamic connectivity estimates obtained from the covariate naive BPMM approach, and subsequently only retaining the covariates with significant associations for analysis using the full model. This approach is expected to work well as long as the screening step does not exclude any important covariates and manages to largely filter out spurious covariates that are unrelated to the network. In future work, we plan to extend the proposed approach to incorporate feature selection that automatically identifies significant covariates that are related to the dynamic networks, and down-weights the contribution of unimportant covariates using Bayesian shrinkage priors.

In addition to identifying important connectivity changes, during the fMRI block task experiment, our analysis conclusively established major changes between the pre- and post-intervention networks that were exclusive to the spin group. We note that existing literature has established the role of cardiovascular fitness in regulating aging related declines in both language and motor control (McGregor et al., 2011, 2013). However, much less is known about the effect of exercise intervention on dynamic connectivity, particularly in older adults. Because connectivity is a fundamental aspect of neuronal communication required for high-level cognitive processes, it is important to understand the potential impact of aging and/or aerobic exercise interventions in aging on changes in brain connectivity.

Further, our analysis also discovered subgroups of individuals with homologous dynamic connectivity, where the heterogeneity within these subgroups with respect to intervention was higher under the idPMAC method compared to the idPAC analysis. This indicates that dynamic pairwise correlations were more accurate in classifying participants in terms of the intervention administered. It is important to note that the separation of clusters with respect to intervention reflects the distinct patterns of dynamic connectivity between the 18 brain regions specified in our study that are known to be differentially activated in spin and control groups (Nocera et al., 2017). However, if additional regions are included that may not be necessarily associated with intervention type, it is entirely possible to obtain more heterogeneous clusters that have a more equitable composition with respect to intervention group. This is due to the presence of network edges between regions that are not necessarily associated with intervention and hence behave similarly in both the spin and control groups. Future work will focus on a more general analysis involving a larger number of cannonical regions known to be associated with the semantic language function.

\section*{Supplementary Materials}
The Supplementary Materials contain additional details corresponding to the M-steps for dynamic pairwise correlations and  partial correlations, as well as details for selecting the tuning parameter in (\ref{eq:multiTV}) for change point estimation corresponding to Section 2.4.

\section*{Acknowledgements}
The views expressed in this work do not necessarily reflect those of the National Institutes of Health, Department of Veterans Affairs or the United States Government. The work was supported by NIMH award number R01MH120299 (SK), and VA research awards: IK2RX000956 (KMM); IK2RX000744 (JN).

\section*{Data and Code Availability}
A portion of the data presented in this work is property of the United States Department of Veterans Affairs. Copies of the de-identified data can be made available upon written request to the corresponding author and Department of Veterans Affairs. The code for implementing the proposed approaches are available here: {\it https://github.com/Emory-CBIS/BPMM}

\section*{Ethics Statement}
Study procedures were approved by the institutional review board of Emory University, informed consent was obtained for experimentation with human subjects, and procedures were consistent with the Declaration of Helsinki.

\section*{Appendix}

\subsection*{Posterior Distribution for Dynamic Pairwise Correlations}

Here, we derive the  log-posterior distribution that is used in the EM algorithm to derive parameter estimates. The augmented log-posterior distribution for $\bm\Theta^{jl}$ under (\ref{eq:base})-(\ref{eq:base_cov}) is:
\begin{align}
&\log(\pi(\bm{\Theta}^{jl}\mid \bm{Y} ) )  \propto  \log \bigg(P(\bm{\Theta}^{jl})P(\bm{Y|\Theta}^{jl}) \bigg) =\sum_{h=1}^H\sum_{t=1}^T\log(\pi(\gamma^*_{h,jlt})) + \sum_{h=1}^H\log(\pi(\sigma^2_{\gamma,h})) \nonumber \\
&+\sum_{h=1}^{H-1}\sum_{t=1}^T\log(\pi(\bfb_{h,jlt})) +\sum_{i=1\dots N} \sum_{t=1\dots T}\log \bigg(P(y^{(i)}_{jt},y^{(i)}_{lt}|\gamma_{jl,t}^{(i)},\sigma_{y}^2) \times \pi(\gamma_{jl,t}^{(i)}) \bigg)  \nonumber \\
 &\propto \sum_{i=1}^N \sum_{t=1}^T \bigg[ -\frac{1}{2}\log\bigg\{1-\bigg(\frac{\exp(2\gamma_{jl,t}^{(i)})-1}{\exp(2\gamma_{jl,t}^{(i)})+1}\bigg)^2\bigg\} - \bigg\{\frac{(y^{(i)}_{jt})^2 + (y^{(i)}_{lt})^2 - 2\big(\frac{\exp(2\gamma_{jl,t}^{(i)})-1}{\exp(2\gamma_{jl,t}^{(i)})+1}\big)y^{(i)}_{jt}y^{(i)}_{lt}}{2\sigma_y^2 \bigg(1-(\frac{\exp(2\gamma_{jl,t}^{(i)})-1}{\exp(2\gamma_{jl,t}^{(i)})+1})^2\bigg)}\bigg\} \nonumber\\
 & -\frac{1}{2}\sum_{h=1}^H\bigg\{\frac{1}{\sigma_{\gamma,h}^2}\Delta^{(i)}_{h,jlt} (\gamma_{jl,t}^{(i)}-\gamma_{h,jlt}^*)^2 + \Delta^{(i)}_{h,jlt}\log(\sigma^2_{\gamma,h})\bigg\}  
+\sum_{h=1}^{H-1}  \Delta_{h,jlt}^{(i)} \bm{x_i}^T \bfb_{h,jlt} - \log \big(1+\sum_{r=1}^{H-1}e^{\bm{x_i}^T \bfb_{r,jlt}}\big) \bigg] \nonumber \\
&+\sum_{h=1}^{H-1}\sum_{t=1}^T\log(\pi(\bfb_{h,jlt}))+ \sum_{h=2}^H \bigg\{\bigg(-\lambda|\gamma^*_{h,jl,t}-\gamma^*_{h,jl,t-1}|\bigg) - (a_{\sigma}+1)\log(\sigma_{\gamma,h}^2) - \frac{b_{\sigma}}{\sigma_{\gamma,h}^2} \bigg\},\label{eq:lpost1}
 \end{align}
 where $\small \log(\pi(\bfb_{h,jlt}))=-\frac{\bfb_{h,jlt}^T\Sigma_{\beta}^{-1}\bfb_{h,jlt}}{2} - \frac{1}{2} \log (det(\Sigma_{\beta})) $ represents the logarithm of the prior distribution on the covariate effects. The detailed computational steps for deriving the MAP estimates corresponding to the above posterior distribution are discussed in Section 3.
 
 \subsection*{Posterior Distribution for Dynamic Precision Matrices}
   The augmented log-posterior distribution for the model parameters can be written as  $\log(\bm{\Theta}\mid Y^{(1)},\ldots,Y^{(N)})$
\begin{align}\small
&\propto \sum_{i=1}^{N}\sum_{t=1}^{T}\log\bigg(P({\bf y}^{(i)}_t\mid \Omega^{(i)}_t) \prod_{v=1}^{V}\pi(\omega^{(i)}_{t,vv}\mid\alpha)\pi({\bm\omega}^{(i)}_{vt}\mid {\bm\omega}_{1,vt}^{*},\ldots,{\bm\omega}_{H,vt}^{*},\sigma^{2}_{\omega,1},\ldots,\sigma^{2}_{\omega,H})\bigg) \nonumber \\
&+\sum_{h=1}^H \sum_{v=1}^H \sum_{t=1}^T \log (\pi(\bm{\omega_{h,vt}^*})) + \sum_{h=1}^H \log(\pi(\sigma_{\omega,h}^2)) \propto \sum_{i=1}^N\sum_{t=1}^T \frac{1}{2} \bigg[ \log \det (\Omega_{11,t}^{(i)}) - {\bf{y}}_{t,-1}^{(i)'} \Omega_{11,t}^{(i)} {\bf{y}}_{t,-1}^{(i)} \bigg ] \nonumber \\
& +\sum_{i=1}^N \sum_{t=1}^T \frac{1}{2} \bigg [ -\log {\kappa}_{1,t}^{(i)} -\big ( s_{11,t}^{(i)}+\alpha \big )\kappa_{1,t}^{(i)} - {\bm{\omega}}_{1t}^{(i)'} \bigg ( \sigma_{\omega,h}^2 I_{V-1} + (s_{11,t}^{(i)}+\alpha) \Omega_{11,t}^{(i)-1} \bigg )  {\bm{\omega}}_{1t}^{(i)} + 2 {\bf s}_{1,t}^{(i)'} {\bm \omega}_{1t}^{(i)} \bigg ] \nonumber \\
 &  -\frac{1}{2}\bigg\{\sum_{h=1}^{H}\sum_{v=1}^{V}\frac{1}{\sigma_{\omega,h}^{2}}\Delta^{(i)}_{h,vt} ({\bm\omega}^{(i)}_{v,t}-{\bm\omega}_{h,t}^{*})'({\bm\omega}^{(i)}_{v,t}-{\bm\omega}_{h,t}^{*}) \bigg\}    -\frac{V(V-1)}{2}\bigg\{\sum_{h=1}^{H} \Delta^{(i)}_{h,vt}\log(\sigma^{2}_{\omega,h})\bigg\} \nonumber \\
&+  \sum_{v=1}^{V}\bigg\{  \sum_{h=1}^{H-1}\Delta^{(i)}_{h,vt} (\bm{x_i}^{T} \bfb_{h,t}) - \log \bigg(1+\sum_{r=1}^{H-1}exp({\bm{x_i}^{T} \bfb_{r,t}}) \bigg) \bigg\}  \bigg] + \sum_{t=1}^{T}\sum_{v=1}^{V}\sum_{h=1}^{H-1}\log(\pi(\bfb_{h,t})) \nonumber \\
&+\sum_{h=1}^{H} \bigg\{\sum_{t=1}^{T}\bigg(-\lambda|{\bm\omega}^{*}_{h,t}-{\bm\omega}^{*}_{h,t-1}|_1\bigg) - (a_{\sigma}+1)\log(\sigma_{\omega,h}^{2}) - \frac{b_{\sigma}}{\sigma_{\omega,h}^{2}} \bigg\}, \label{eq:lpost2}
\end{align}
where  $\small \log(\pi(\bfb_{h,t}))=-\frac{\bfb_{h,t}^T\Sigma_{\beta}^{-1}\bfb_{h,t}}{2} - \frac{1}{2} \log (det(\Sigma_{\beta})) $ represents the logarithm of the prior distribution on the covariate effects. The EM algorithm to derive the MAP estimators for model parameters is based on the expression for the above log-posterior (see Section 3).

\subsection*{M-steps for dynamic pairwise correlations}

{\noindent \bf M-step for mixture atoms:}
Denote $\bm{\gamma}_{h,jl}=(\gamma_{h,jl,1},\ldots,\gamma_{h,jl,T})$, $\bar{\gamma}_{h,jl,t}= \frac{1}{\sum_{i=1}^N \hat{\psi}_{h,jlt}^{(i)}}\sum_{i=1}^N \hat{\psi}_{h,jlt}^{(i)}\gamma_{jl,t}^{(i)}$, $w_{h,jlt}=\frac{\sum_{i=1}^N\hat{\psi}_{h,jlt}^{(i)}}{2\sigma^2_{\gamma,h}}$, and  $\bar{\gamma}^{(w)}_{h,jl,t}=\sqrt{w_{h,jlt}}\bar{\gamma}_{h,jl,t}$. Further denote $|\cdot|$ as the element-wise $L_1$ norm, and denote  $\bm{\tilde{\eta}}_{h,jl}=(\tilde{\eta}_{h,jl,0},\tilde{\eta}_{h,jl,1},\ldots, \tilde{\eta}_{h,jl,T-1})$, $\tilde{\eta}_{h,jl,0}=\gamma^*_{h,jl,1}$,  $\tilde{\eta}_{hjl,t-1}=\gamma^*_{h,jl,t}-\gamma^*_{h,jl,t-1}$. Then, using the derivations presented in the Supplementary Materials,
$
\small
\widehat{\bm{\gamma}}^*_{h,jl} = \arg\min\sum_{t=1}^T (\sqrt{w_{h,jlt}}\bar{\gamma}_{h,jl,t} - \sqrt{w_{h,jlt}}\gamma^*_{h,jl,t})^2 + \lambda\sum_{t=1}^{T-1}|\gamma^*_{h,jlt}-\gamma^*_{h,jl,t-1}|
= \arg\min|| \bar{\bm{\gamma}}^{(w)}_{h,jl}  - \tilde{M}_{h,jl}\bm{\tilde{\eta}}_{h,jl}||^2 + \lambda\sum_{t=0}^{T-1}|\tilde{\eta}_{h,jl,t}|, 
$
where the $T\times T$ matrix $\tilde{M}_{h,jl}$ has the following form 
\begin{eqnarray*}
\tilde{M}_{h,jl}=
\begin{bmatrix}
\sqrt{w_{h,jl,1}} &0 &0 \ldots &0\\
\sqrt{w_{h,jl,2}} &\sqrt{w_{h,jl,2}} &0 \ldots &0 \\
\sqrt{w_{h,jl,3}} &\sqrt{w_{h,jl,3}} &0 \ldots &0\\
\vdots            &\vdots            &\vdots &\vdots\\
\sqrt{w_{h,jl,T}} &\sqrt{w_{h,jl,T}} & &\sqrt{w_{h,jl,T}}
\end{bmatrix}.
\end{eqnarray*}
The solution can be obtained using a Lasso algorithm with the penalty parameter $\lambda$ being chosen using BIC. The solutions for $\bm{\tilde{\eta}}_{h,jl}$ can be directly used to recover the estimates for ${\bm\gamma}^*_{h,jl}=(\gamma^*_{h,jl,1},\ldots,\gamma^*_{h,jl,T} )$, which in turn yields the dynamic connectivity estimates.

{\noindent \bf  M-step for mixture variance:}  
Use the closed form solution to estimate ($h=1,\ldots,H$):\\
$\hat{\sigma_{\gamma,h}^2} = \bigg(a_{\sigma}+0.5\sum_{t=1}^T\sum_{i=1}^N \hat{\psi}_{h,jlt}^{(i)}-1\bigg)^{-1}\bigg(b_{\sigma}+ 0.5\sum_{t=1}^T\sum_{i=1}^N \hat{\psi}_{h,jlt}^{(i)}(\gamma_{jl,t}^{(i)}-\hat{\gamma}_{h,jlt}^*)^2)\bigg)$.



{\noindent \bf  M-step for pair-wise correlations:}
The update of $\gamma_{jl,t}^{(i)}$ is performed via a  Newton-Raphson step. Denote the parameter estimate at $f$-th iteration of Newton-Raphson as $\gamma_{jl,t}^{(i)[f]}$, and use the update for the $(f+1)$-th iteration as 
 $\gamma_{jl,t}^{(i)[f+1]} = \gamma_{jl,t}^{(i)[f]} - \frac{a_1(\gamma_{jl,t}^{(i)[f]})}{a_2(\gamma_{jl,t}^{(i)[f]})}$, where $a_1(\gamma_{jl,t}^{(i)[f]})$ and $a_1(\gamma_{jl,t}^{(i)[f]})$ are expressed as:
\begin{align*}
a_1(\gamma_{jl,t}^{(i)[f]}) &= d(i(\Theta)^{[f]})/d(\gamma^{(i)[f]}_{jl,t}) = \frac{exp(2\gamma_{jl,t}^{(i)[f]})-1}{exp(2\gamma_{jl,t}^{(i)[f]})+1} - \sum_{h=1}^H\frac{\hat{\psi}_{h,jlt}^{(i)}(\gamma_{jl,t}^{(i)[f]}-\gamma_{h,jlt}^*)}{\sigma_{\gamma,h,jlt}^2 }\\
&- \frac{(exp(2\gamma_{jl,t}^{(i)[f]})^2-1)(y_{jt}^2+y_{lt}^2)-2y_{jt}y_{lt}(exp(2\gamma_{jl,t}^{(i)[f]})^2+1)}{4\sigma_y^2 exp(2\gamma_{jl,t}^{(i)[f]})}, \mbox{ and } \\
   a_2(\gamma_{jl,t}^{(i)[f]}) &= d(i(\Theta)^{[f]2})/d(\gamma^{(i)[f]}_{jl,t})^2= \frac{4exp(2\gamma^{(i)[f]}_{jl,t})}{(exp(2\gamma^{(i)[f]}_{jl,t})+1)^2} -\sum_{h=1}^H \frac{\hat{\psi}_{h,jlt}^{(i)}}{\sigma_{\gamma,h,jlt}^2} \\
   & -\frac{exp(2\gamma^{(i)[f]}_{jl,t})^2(y_{jt}^2+y_{lt}^2-2y_{jt}y_{lt})+(y_{jt}^2+y_{lt}^2+y_{jt}y_{lt})}{2\sigma_y^2exp(2\gamma^{(i)[f]}_{jl,t})} 
\end{align*}
The above iterative steps are repeated until convergence, i.e. when $|{\gamma_{jl,t}^{(i)[f+1]} - \gamma_{jl,t}^{(i)[f]}|} < 10^{-3}$.

{\noindent \bf M-step for covariate effects:}
The log-posterior $\log(\pi(\bfb_{h,jlt}\mid -))\propto$ 
 \begin{eqnarray*} \label{eq:logbeta1}\small
 && \bigg\{-\frac{\bfb_{h,jlt}^T\Sigma_{\beta}^{-1}\bfb_{h,jlt}}{2} - \frac{1}{2} \log (det(\Sigma_{\beta})) \bigg\} 
 + \sum_{i=1}^N  \bigg\{ \Delta_{h,jlt}^{(i)} \bm{x_i}^T \bfb_{h,jlt} - 
 \log \big(1+\sum_{r=1}^{H-1}exp(\bm{x_i}^T \bfb_{r,jlt})\big)
\bigg\} \\
&&\approx  -\frac{1}{2}\sum_{i=1}^N w_{h,jlt}(z_{h,jlt} - \bm{x_i}^T \bfb_{h,jlt})^2 -\frac{\bfb_{h,jlt}^T\Sigma_{\beta}^{-1}\bfb_{h,jlt}}{2},
\end{eqnarray*}
using the expression in (\ref{eq:lpost1}), and a quadratic approximation as in (Friedman et al., 2010) for the last step, in order to facilitate closed form updates. In the above expression, $ \small z_{h,jlt} = \bm{x_i}^T \tilde{\bfb_{h,jlt}} + \frac{\widehat{\Delta}_{h,jlt}^{(i)}-\tilde{p}_{h,jlt}({\bm x_i})}{\tilde{p}_{h,jlt}({\bm x_i})(1-\tilde{p}_{h,jlt}({\bm x_i}))}$, $ w_{h,jlt}=\tilde{p}_{h,jlt}(\bm_{x_i})(1-\tilde{p}_{h,jlt}({\bm x_i}))$,  $\tilde{p}_{h,jlt} = \tilde{P}(\Delta^{(i)}_{h,jlt} = 1 \mid {{\bm x}_i})=\frac{exp(\bm{x_i}^T \check{\bfb}_{h,jlt})}{1 + \sum_{h=1}^{H-1}exp(\bm{x_i}^T\check{\bfb}_{h,jlt})}$ represents the approximated probability under the quadratic approximation, $\check{\bfb}_{h,jlt}$ represents the estimate of $\bfb_{h,jlt}$ at previous step, and $\widehat{\Delta}_{h,jlt}^{(i)}$ represents expected probability for the $i$-th subject as in the E-step. The above approximate log-posterior can be optimized to obtain a closed form expression as 
$\widehat{\bfb}_{h,jlt}=\arg\max_{\bfb}   \log(\pi(\bfb_{h,jlt}\mid - )) 
  = (\Sigma_{\beta}^{-1}+\sum_{i=1}^Nw_{h,jl}\bm{x_i}\bm{x_i}^T)^{-1}(\sum_{i=1}^N w_{h,jl}z_{h,jl}\bm{x_i}),$
where the notations in the expression for $\widehat{\bfb}_{h,jlt}$ has been defined previously.

\subsection*{M-steps for dynamic precision matrix estimation}

{\noindent \bf M-step for mixture atoms:} Define ${\bf e}^*_{h,t}=({\bm\omega}^*_{h,t}-{\bm\omega}^*_{h,t-1})'=(e^*_{h,1t},\ldots,e^*_{h,Vt}),t=1,\ldots,T-1$, $E^*_{h}=({\bm\omega}^*_{h}, {\bf e}^*_{h,1},\ldots,{\bf e}^*_{h,T-1})'$, $\{e^*_{h,v't} \}$ represents the elements in ${\bf e}^*_{h,t}, \bar{W}_{h,v}$ is a $T\times V-1$ matrix with the $t$-th row as $\frac{\sum_{i=1}^{N} \Delta_{h,vt}^{(i))} \bm{\omega}_{v,t}^{(i)}}{2\sigma_{\omega,h}^2}$ ,
 $\bar{W}_{h,v}(\bullet, v')$ and $E^*_{h}(\bullet, v')$ represent the $v'$th column of $\bar{W}_{h,v}$ and $E^*_{h}$ respectively,  and 
$|\cdot|_1$ represents element-wise $L_1$ norm. Similar to the steps for dynamic pairwise correlations, the estimate for mixture atom ${\bm\omega}^*_{h,t}, h=1,\ldots,H, t=1,\ldots,T,$ can be obtained by minimizing the following objective function:
\begin{eqnarray*}
\small
 \sum_{v=1}^V \{|| \bar{W}_{h,v} - M^*_{h,v}E^*_{h}||^2 &+& \lambda\sum_{t=1}^{T-1}|{\bf e}^*_{h,t}|_1 \}
=  \sum_{v=1}^V \sum_{v'=1}^{V-1} \{|| \bar{W}_{h,v}(\bullet, v') - M^*_{h,v}E^*_{h}(\bullet, v')||^2 + \lambda\sum_{t=1}^{T-1}|e^*_{h,v't}|_1  \}, \\
\mbox{ where } M^*_{h,v} &=&
\begin{bmatrix}
\sqrt{w_{h,v,1}} &0 &0 \ldots &0\\
\sqrt{w_{h,v,2}} &\sqrt{w_{h,v,2}} &0 \ldots &0 \\
\sqrt{w_{h,v,3}} &\sqrt{w_{h,v,3}} &\sqrt{w_{h,v,3}} \ldots &0\\
\vdots            &\vdots            &\vdots &\vdots\\
\sqrt{w_{h,v,T}} &\sqrt{w_{h,v,T}}  &\sqrt{w_{h,v,T}} \ldots &\sqrt{w_{h,v,T}} 
\end{bmatrix}.
\end{eqnarray*}
The above equation can be solved using a Lasso algorithm with the penalty parameter $\lambda$ being chosen using BIC. The solutions for $\bm{E}^*_{h}$ are then used to recover the estimates for ${\bf \omega}^*_{h,t}$. 

{\noindent \bf  M-step for mixture variance:} Use 
$\hat{\sigma}^2_{\omega,h}= \frac{b_\sigma + 0.5\sum_{i=1}^N\sum_{t=1}^T\sum_{v=1}^V \Delta^{(i)}_{h,vt} ({\bm\omega}^{(i)}_{v,t}-{\bm\omega}_{h,t}^*)'({\bm\omega}^{(i)}_{v,t}-{\bm\omega}_{h,t}^*)}{a_\sigma+1+0.5V(V-1)\sum_{t=1}^T\sum_{i=1}^N \Delta^{(i)}_{h,vt}}$.

{\noindent \bf M-step for covariate effects:}
Using similar arguments as in Section 3.1, one can approximate the posterior as: 
\begin{align*}
\small
   \log(\pi(\bfb_{h,t}\mid - )) \approx  -\frac{1}{2}\sum_{i=1}^N\sum_{v=1}^V w_{h,t}(z_{h,vt} - \bm{x_i}^T \bfb_{h,t})^2 -\frac{\bfb_{h,t}^T\Sigma_{\beta}^{-1}\bfb_{h,t}}{2},
  \end{align*}
where $ z_{h,vt} = \bm{x_i}^T \tilde{\bfb_{h,t}} + \frac{\widehat{\psi}_{h,vt}^{(i)}-\tilde{p}_{h,t}({\bm x_i})}{\tilde{p}_{h,t}({\bm x_i})(1-\tilde{p}_{h,t}({\bm x_i}))}$, $ w_{h,t}=\tilde{p}_{h,t}({\bf x}_i)(1-\tilde{p}_{h,t}({\bf x}_i))$, $\tilde{p}_{h,t} = \tilde{P}(\Delta^{(i)}_{h,t} = 1 \mid {{\bm x}_i})=\frac{exp(\bm{x_i}^T \tilde{\bfb}_{h,t})}{1 + \sum_{h=1}^{H-1}exp(\bm{x_i}^T\tilde{\bfb}_{h,t})}$ represents the approximated probability under the quadratic approximation, where $\tilde{\bfb}_{h,t}$ denotes the estimate of $\bfb_{h,t}$ at previous step, and $\widehat{\psi}_{h,vt}^{(i)}$ represents expected probability for subject $i$ as calculated in the E-step. The above approximate log-likelihood can be optimized to obtain a closed form expression $
 \widehat{\bfb}_{h,t}= (\Sigma_{\beta}^{-1}+ V\sum_{i=1}^N w_{h,t}\bm{x_i}\bm{x_i}^T)^{-1}(\sum_{i=1}^N\sum_{v=1}^V w_{h,t}z_{h,vt}\bm{x_i}),$
where the notations in the expression for $\widehat{\bfb}_{h,vt}$ has been defined previously.

\section*{References}

\begin{enumerate}
\item Allen, E. A., Damaraju, E., Plis, S. M., Erhardt, E. B., Eichele, T., and Calhoun, V. D. (2014). Tracking whole-brain connectivity dynamics in the resting state. Cerebral cortex, 24(3), 663-676.
\item Balsters, J. H., Whelan, C. D., Robertson, I. H., and Ramnani, N. (2013). Cerebellum and cognition: evidence for the encoding of higher order rules. Cerebral Cortex, 23(6), 1433-1443.
\item Becker, R. A., Chambers, J. M., and Wilks, A. R. (1988). The New S Language. Wadsworth \& Brooks. Cole.[Google Scholar].
\item Bullmore, E., and Sporns, O. (2009). Complex brain networks: graph theoretical analysis of structural and functional systems. Nature reviews neuroscience, 10(3), 186-198.
\item Chang, C., \& Glover, G. H. (2010). Time–frequency dynamics of resting-state brain connectivity measured with fMRI. Neuroimage, 50(1), 81-98.
\item Cribben, I., Wager, T., and Lindquist, M. (2013). Detecting functional connectivity change points for single-subject fMRI data. Frontiers in computational neuroscience, 7, 143.
\item Davey, J., Thompson, H. E., Hallam, G., Karapanagiotidis, T., Murphy, C., De Caso, I., ... and Jefferies, E. (2016). Exploring the role of the posterior middle temporal gyrus in semantic cognition: Integration of anterior temporal lobe with executive processes. Neuroimage, 137, 165-177.
\item D'Mello AM, Turkeltaub PE, and Stoodley CJ. (2017). Cerebellar tDCS Modulates Neural Circuits during Semantic Prediction: A Combined tDCS-fMRI Study. J Neuroscience;37(6):1604-1613.
\item Durante, D., Dunson, D. B., and Vogelstein, J. T. (2017), “Nonparametric Bayes modeling of populations of networks,” Journal of the American Statistical Association, 112, 1516–1530.
\item Engel, J. (1988), Polytomous logistic regression. Statistica Neerlandica, 42: 233-252. 
\item Filippi, M., Spinelli, E. G., Cividini, C., and Agosta, F. (2019). Resting state dynamic functional connectivity in neurodegenerative conditions: a review of magnetic resonance imaging findings. Frontiers in neuroscience, 13, 657.
\item Hidot, S., and Saint-Jean, C. (2010). An Expectation–Maximization algorithm for the Wishart mixture model: Application to movement clustering. Pattern Recognition Letters, 31(14), 2318-2324.
\item Hindriks, R., Adhikari, M. H., Murayama, Y., Ganzetti, M., Mantini, D., Logothetis, N. K., \& Deco, G. (2016). Can sliding-window correlations reveal dynamic functional connectivity in resting-state fMRI?. Neuroimage, 127, 242-256.
\item Hutchison, R. M., Womelsdorf, T., Allen, E. A., Bandettini, P. A., Calhoun, V. D., Corbetta, M., ... and Handwerker, D. A. (2013). Dynamic functional connectivity: promise, issues, and interpretations. Neuroimage, 80, 360-378.
\item Jolles, D. D., van Buchem, M. A., Crone, E. A., \& Rombouts, S. A. (2013). Functional brain connectivity at rest changes after working memory training. Human brain mapping, 34(2), 396-406.
\item Kundu, S., Ming, J., Pierce, J., McDowell, J., \& Guo, Y. (2018). Estimating dynamic brain functional networks using multi-subject fMRI data. NeuroImage, 183, 635-649.
\item Lacombe, J., Jolicoeur, P., Grimault, S., Pineault, J., and Joubert, S. (2015). Neural changes associated with semantic processing in healthy aging despite intact behavioral performance. Brain and language, 149, 118-127.
\item Laufer, I., Negishi, M., Lacadie, C. M., Papademetris, X., and Constable, R. T. (2011). Dissociation between the activity of the right middle frontal gyrus and the middle temporal gyrus in processing semantic priming. PloS one, 6(8), e22368.
\item Lindquist, M. A., Xu, Y., Nebel, M. B., \& Caffo, B. S. (2014). Evaluating dynamic bivariate correlations in resting-state fMRI: a comparison study and a new approach. NeuroImage, 101, 531-546.
\item Lukemire, J., Kundu, S., Pagnoni, G., \& Guo, Y. (2020). Bayesian joint modeling of multiple brain functional networks. Journal of the American Statistical Association, 1-13.
\item MacEachern, S. N. (1999). Dependent nonparametric processes. In ASA Proceedings of the Section on Bayesian Statistical Science, Alexandria, VA. American Statistical Association.
\item McGregor, K. M., Zlatar, Z., Kleim, E., Sudhyadhom, A., Bauer, A., Phan, S., et al. (2011). Physical activity and neural correlates of aging: a combined TMS/fMRI study. Behav. Brain Res. 222, 158–168.
\item Meil$\check{a}$, M. (2007). Comparing  —an information based distance. Journal of multivariate analysis, 98(5), 873-895.
\item Moberget, T., Gullesen, E. H., Andersson, S., Ivry, R. B., and Endestad, T. (2014). Generalized role for the cerebellum in encoding internal models: evidence from semantic processing. The Journal of neuroscience : the official journal of the Society for Neuroscience, 34(8), 2871–2878. 
\item Monti, R. P., Hellyer, P., Sharp, D., Leech, R., Anagnostopoulos, C., \& Montana, G. (2014). Estimating time-varying brain connectivity networks from functional MRI time series. NeuroImage, 103, 427-443.
\item Nielsen, S. F. V., Madsen, K. H., Schmidt, M. N., and Mørup, M. (2017). Modeling dynamic functional connectivity using a wishart mixture model. In Proceedings of the 2017 International Workshop on Pattern Recognition in Neuroimaging (pp. 1-4). IEEE. 2017 International Workshop on Pattern Recognition in Neuroimaging (prni) https://doi.org/10.1109/PRNI.2017.7981505
\item Nocera, J., Crosson, B., Mammino, K., and McGregor, K. M. (2017). Changes in cortical activation patterns in language areas following an aerobic exercise intervention in older adults. Neural Plasticity, 2017.
\item Patrikainen A. and Meila M. (2006). Comparing subspace clusterings. IEEE Transactions on Knowledge and Data Engineering 18, 902–916.
\item Quinn, A. J., Vidaurre, D., Abeysuriya, R., Becker, R., Nobre, A. C., and Woolrich, M. W. (2018). Task-evoked dynamic network analysis through hidden markov modeling. Frontiers in neuroscience, 12, 603.
\item Shi, R., and Guo, Y. (2016). Investigating differences in brain functional networks using hierarchical covariate-adjusted independent component analysis. The annals of applied statistics, 10(4), 1930.
\item Smith, S. M., Beckmann, C. F., Andersson, J., Auerbach, E. J., Bijsterbosch, J., Douaud, G., ... \& Kelly, M. (2013). Resting-state fMRI in the human connectome project. Neuroimage, 80, 144-168.
\item Sun, W. W., and Li, L. (2017). STORE: sparse tensor response regression and neuroimaging analysis. The Journal of Machine Learning Research, 18(1), 4908-4944.
\item Thorndike, R. L. (1953). Who belongs in the family?. Psychometrika, 18(4), 267-276.
\item Tibshirani, R., \& Wang, P. (2008). Spatial smoothing and hot spot detection for CGH data using the fused lasso. Biostatistics, 9(1), 18-29.
\item Vert, J. P., \& Bleakley, K. (2010). Fast detection of multiple change-points shared by many signals using group LARS. In Advances in neural information processing systems (pp. 2343-2351).
\item Wang, H. (2012). Bayesian graphical lasso models and efficient posterior computation. Bayesian Analysis, 7(4), 867-886.
\item Wang H, He W, Wu J, Zhang J, Jin Z, and Li L. A coordinate-based meta-analysis of the n-back working memory paradigm using activation likelihood estimation. Brain Cogn. 2019 Jun;132:1-12.
\item Wang, L., Zhang, Z., and Dunson, D. (2019). Common and individual structure of brain networks. The Annals of Applied Statistics, 13(1), 85-112.
\item Warnick, R., Guindani, M., Erhardt, E., Allen, E., Calhoun, V., and Vannucci, M. (2018). A Bayesian approach for estimating dynamic functional network connectivity in fMRI data. Journal of the American Statistical Association, 113(521), 134-151.
\item Wei, G. C. G. and Tanner, M. A. (1990). A Monte Carlo implementation of the EM algorithm and the poor man’s data augmentation algorithms. Journal of the American Statistical Association 85 699–704.
\item Zhang, J., Sun, W. W., and Li, L. (2018). Network response regression for modeling population of networks with covariates. arXiv preprint arXiv:1810.03192.
\item Zhang, Z., Allen, G. I., Zhu, H., and Dunson, D. (2019). Tensor network factorizations: Relationships between brain structural connectomes and traits. Neuroimage, 197, 330-343.
\end{enumerate}

\end{document}